\documentclass[a4paper,11pt]{article}
\pdfoutput=1
\usepackage[UKenglish]{babel}
\usepackage{jheppub}
\usepackage{cancel,tensor,enumitem,fix-cm}
\usepackage{epsfig}
\usepackage{graphicx}
\usepackage{amssymb}
\usepackage{dsfont}
\usepackage[utf8]{inputenc}
\usepackage{textcomp}
\usepackage{multirow}
\usepackage{booktabs}
\usepackage{tikz}
\usepackage{overpic}
\usepackage{hyperref}
\usepackage{amsthm}
\usepackage{amsmath}     
\usepackage[normalem]{ulem}
\usepackage{makecell}
  {
      \theoremstyle{plain}
      \newtheorem{assumption}{Assumption}
  }
\usepackage{multirow}  
\newcommand\scalemath[2]{\scalebox{#1}{\mbox{\ensuremath{\displaystyle #2}}}}
\newcommand\cqstate{\varrho}

\newcommand\lapse{{\cal H}}

\newcommand\g{{g}}

\newcommand\ch{ \mathcal{H} }
\newcommand\qh{ \mathcal{H}_m}
\newcommand\cm{ \mathcal{H} }
\newcommand\qm{ \mathcal{H}}
\newcommand\bphi{ {\bf \phi}}
\newcommand\bpi{ {\bf \pi}}
\newcommand\parts{\mathcal{R}}
\newcommand\con{C}
\newcommand\gauge{[N,\vec{N}]}
\newcommand{\lin}{L}
\newcommand{\ag}{{\boldsymbol\alpha}}
\newcommand{\bg}{{\boldsymbol\beta}}
\newcommand\s{\,\,\,\,}
\newcommand{\Tr}{\mathop{\mathsf{Tr}}\nolimits}

\begin{document}

\title{The constraints of post-quantum classical gravity}
\author[a]{Jonathan Oppenheim}
\author[a]{ Zachary Weller-Davies}
\affiliation[a]{Department of Physics and Astronomy, University College London, Gower Street, London WC1E 6BT, United Kingdom}
\emailAdd{j.oppenheim@ucl.ac.uk}
\emailAdd{zachary.weller-davies@ucl.ac.uk}

\abstract{
We study a class of theories in which space-time is treated classically, while interacting with quantum fields. These circumvent various no-go theorems and the pathologies of semi-classical gravity, by being linear in the density matrix and phase-space density. The theory can either be considered fundamental or as an effective theory where the classical limit is taken of space-time. The theories have the dynamics of general relativity as their classical limit and provide a way to study the back-action of quantum fields on the space-time metric. The theory is invariant under spatial diffeomorphisms, and here, we provide a methodology to derive the constraint equations of such a theory by imposing invariance of the dynamics under  time-reparametrization invariance. This leads to generalisations of the Hamiltonian and momentum constraints. We compute the constraint algebra for a wide class of realisations of the theory (the ``discrete class'') in the case of a quantum scalar field interacting with gravity. We find that the algebra doesn't close without additional constraints, although these do not necessarily reduce the number of local degrees of freedom.
}
\maketitle

\section{Introduction}\label{sec: Intro}
We are often interested in studying the back-reaction of quantum fields on space-time, in the limit where space-time can be treated classically. Current methods to do this, such as the semi-classical Einstein's equations~\cite{moller1962theories,rosenfeld1963quantization}, are only valid when quantum fluctuations are small. Recent experimental proposals~\cite{bose2017spin,marletto2017gravitationally,decodiff2}, have also re-ignited discussion on whether one can fundamentally treat space-time classically.
The question of whether or not gravity is quantum-mechanical is, ultimately, one for experiment. The problem is simple, in the presence of quantum matter, the Einstein field equations
\begin{equation}
G_{\mu \nu}  \stackrel{?}{=} \frac{8 \pi G}{c^4}\hat{T}_{\mu \nu}
\label{eq: Ein}
\end{equation}no longer make sense since the right hand side is an operator and the left hand side a c-number. There are two natural resolutions: either you try and quantize gravity, or else you try and treat it classically by modifying Einstein's equations in some way. Simple interpretations of the later fail; one cannot treat \eqref{eq: Ein} as an eigenvalue equation since the components of the stress-energy tensor fail to commute, whilst the semi-classical equations, obtained by replacing $T_{\mu \nu}$ by its expectation value, $\langle \psi |T_{\mu \nu} | \psi \rangle$, gives rise to non-linearity's, as well as pathological behaviour when the quantum state is in super-position or even a statistical mixture of more than one configuration~\cite{page1981indirect,eppley1977necessity}. 

There are other no-go theorems and arguments concerning the consistency of theories which contain interactions between classical and quantum systems (sometimes called CQ, or hybrid theories). Feynman famously argued that such a coupling would prevent superpositions~\cite{cecile2011role}, Epply and Hannah argued that it would lead to super-luminal signalling~\cite{eppley1977necessity}, and various other arguments have been offered over the years~\cite{bohr1933question, dewitt1962definition,caro1999impediments,salcedo1996absence,
sahoo2004mixing,terno2006inconsistency,salcedo2012statistical,barcelo2012hybrid,marletto2017we}. This has led to the widespread belief that gravity must be quantized. Despite this, various attempts have been made to study a classical theory of gravity interacting with quantum systems, for example by using a channel or measurement based approach~\cite{kafri2014classical,tilloy2016sourcing}, or master equations which go by the name of hybrid dynamics but which may not be completely positive \cite{boucher1988semiclassical,hall2005interacting}. However, it is now known that these no-go theorems can be circumvented by allowing for stochastic coupling between the classical and quantum degrees of freedom. Consistent completely positive dynamics which couples classical and quantum degrees of freedom was introduced in~\cite{blanchard1995event, diosi1995quantum} with the CQ-state of the system being represented as a distribution over the classical degrees of freedom, and a subnormalised density matrix at each phase space point living in Hilbert space. The dynamics is linear in the CQ-state, preserves it's form, and is completely positive and normalisation preserving. This  is a necessary and sufficient for the CQ-state to give positive probability outcomes for measurements. This consistent master equation has since been studied in a number of contexts~\cite{diosi2014hybrid,alicki2003completely}, including models of Newtonian gravity~\cite{diosi2011gravity},%,tilloy2016sourcing}
 field theory models of information loss in black holes \cite{poulinKITP}, and General Relativity coupled to quantum field theory \cite{oppenheim_post-quantum_2018}. 

 This now leads to a new line of research: whether fundamental or effective, it is now possible to have a theory of gravity, where one treats the gravitational degrees of freedom as classical and matter as quantum. In \cite{oppenheim_post-quantum_2018} it was shown that one can construct CQ dynamics which reproduces general Hamiltonian dynamics in the classical limit, and in particular, the dynamics of general relativity. Quantum mechanics is modified \cite{diosi1995quantum,tilloy2016sourcing,poulinKITP,oppenheim_post-quantum_2018} -- one necessarily \cite{decodiff2} has information loss and the classical-quantum interaction causes decoherence of the quantum  fields living on space-time. We thus refer to it as a post-quantum theory. With the dynamics of general relativity successfully reproduced, we are now interested in  ensuring that the theory has the appropriate gauge invariance. In particular, Einstein's theory of general relativity is diffeomorphism invariant, and in the Hamiltonian formulation, this is imposed by ensuring that the constraint equations are satisfied. It is here, that other attempts to reconcile quantum theory with gravity have failed. In loop quantum gravity, the constraint algebra doesn't close off-shell~\cite{constraintThiemann:2006cf,Nicolai_2005}, although it is hoped that one can find an operator ordering such that it will. String theory is background dependent, although it is hoped that string field theory will resolve this. 

In the ADM formulation of classical gravity~\cite{arnowitt2008republication,dewitt1967quantum}, the dynamics is generated by a Hamiltonian $H_{ADM} = \int d^3 x N(x) \mathcal{H}(x) + N^a(x) \mathcal{H}_a(x)$ containing freely chosen ``lapse'' $N(x)$, and  ``shift" $N^a(x)$ functions. In order to ensure that the dynamics doesn't depend on this choice, one must impose $\mathcal{H}\approx 0$, $ \mathcal{H}_a \approx 0$, called the Hamiltonian and momentum constraint respectively. The notation $\approx$ is used to indicate that the constraints are {\it weakly zero}, meaning they only vanish on part of the phase space, called the constraint surface.  We here  begin the study of constraints in CQ theories of gravity. In particular, we study a class of classical-quantum theories of gravity which retain some desirable properties, namely local time reparameterization invariance and spatial diffeomorphism invariance. In such models, by applying a version of the Dirac argument (which in its original form relates terms appearing with free Lagrange multipliers, i.e, first class constraints,  to equal time gauge generators) we are able to derive a set of constraints which must be satisfied in order for the theory to be independent of the choice of lapse and shift. In a purely classical theory, one can restrict phase space variables to lie in the constraint surface. In a quantum theory, one can impose the constraints as projectors onto a subspace of the Hilbert space. For a theory which combines classical and quantum degrees of freedom, a different method is required, especially since it is not, as far as we know, easily derived from an action principle which would make the required gauge invariance manifest. Instead, gauge freedom must be imposed on the equations of motion.  We find that independence of the choice of lapse and shift require ``commutation'' (in a certain sense) of the spatial and temporal parts of the equations of motion. For this commutation to hold, one finds a condition on the CQ-state, analogous to the momentum constraint. Conservation of this condition with time, leads to the Hamiltonian constraint. This methodology is the central result of the paper.

We study these constraints explicitly in a theory of a quantum scalar field coupled to classical gravity. Although we are not able to construct a realisation of the theory which satisfies the constraint conditions, at least without a further restriction on the choice of lapse and shift, we expect that finding the correct gauge conditions which have GR in the appropriate limit will allow such a construction.  We also expect the general procedure for deriving the constraints will enable further study  CQ theories of gravity in a concrete setting. We also find that a number of conceptual issues which appear for example, in other attempts to quantize gravity, have an analogous yet simpler form in these theories, where the wave function corresponding the gravitational degrees of freedom is replaced by a probability density. 

The outline of this paper is as follows, in section \ref{sec: CQtheory} we review the formalism introduced in \cite{oppenheim_post-quantum_2018}, introducing CQ dynamics and CQ dynamics which becomes Hamiltonian in the classical limit. In section \ref{sec: CQeinstein}, we introduce the post-quantum theories of gravity which we study in this work; namely those linear in the lapse and shift. In section \ref{sec: derivingconstraints} we first review the Dirac argument, showing in \ref{sec: constraintdirac} how one can use the Dirac argument to derive constraints in classical gravity, before extending the Dirac argument to the case of the CQ master equations in section \ref{sec: diracmaster}. In section \ref{sec: constraintfull} we apply this formalism to theories of gravity, first showing in \ref{sec: genmethod} how one will generically get constraints in theories linear in the lapse and shift. The remainder of the paper is then spent studying the constraints and their preservation in time, in a specific subset of possible realisations of a quantum scalar field interacting with classical gravitational fields in section \ref{sec: scalarfield}. Our results suggest that either a wider class of realisations needs to be considered as additional constraints need to be added for the constraint algebroid to close or alternatively, that one should consider a smaller gauge group by placing some restrictions on the choice of lapse and shift. We conclude our findings with a discussion in section \ref{sec: discussion}. 

%%%%%%%%%%%%%%%%%%%%%%%%%%%%%%%%%%%
\section{CQ theory }\label{sec: CQtheory}
In this section we review classical-quantum dynamics and introduce the formalism which provides a basis for the rest of the paper. For a more detailed discussion on CQ dynamics we refer the reader to \cite{oppenheim_post-quantum_2018, diosi2014hybrid}. We shall first introduce CQ dynamics in its full generality, before focusing on CQ dynamics which reproduces Hamiltonian evolution in the classical limit \cite{oppenheim_post-quantum_2018}. We will sketch the derivation of the general form of the master-equation, but the reader can skip directly to it, at Equation
\eqref{eq: expansion} and its simpler form Equation \eqref{eq: expansion2}.

\subsection{Classical-quantum dynamics}\label{sec: CQdynamics}
We first review the most general master equation with bounded operators governing CQ dynamics. We shall take the classical degrees of freedom to live in a phase space $\Gamma=(\mathcal{M}, \omega)$ and we generically denote elements of the classical space by $z$. For example, we could take the classical degrees of freedom to be position and momenta in which case $\Gamma= (\mathbb{R}^{2}, \omega)$, where $\omega$ is the symplectic form in classical phase space and $z= (q,p)$. In this paper, we shall be interested in the case where the classical degrees of freedom are taken to be Riemannian 3-metrics $g_{ab}$ on a 3-surface $\Sigma$ and their conjugate momenta $\pi^{ab}$ -- so that $z= (g_{ab}, \pi^{ab})$. We shall discuss the details of the gravitational case in section \ref{sec: CQeinstein}. The quantum degrees of freedom are described by a Hilbert space $\mathcal{H}$. Given the Hilbert space, we denote the set of positive semi-definite operators with trace at most unity as $S_{\leq 1}(\mathcal{H})$. Then the CQ object defining the state of the hybrid system at a given time is a map $\cqstate : \mathcal{M} \to  S_{\leq 1}(\mathcal{H})$ subject to a normalization constraint $\int_{\mathcal{M}} dz \Tr{\cqstate} =1$. To put it differently, we associate to each classical degree of freedom a sub-normalized density operator, $\cqstate(z)$, such that $\Tr{\cqstate} = p(z) \geq 0$ is a normalized probability distribution over the classical degrees of freedom and $\int_{\mathcal{M}} dz \cqstate(z) $ is a normalized density operator on $\mathcal{H}$. An example of such a {\it CQ-state} is the CQ qubit, where we take a 2 dimensional Hilbert space and couple to classical position and momenta. The state then takes the form of a  $2\times 2$ matrix over phase space 
\begin{equation}
\cqstate(q, p, t)=\left(\begin{array}{ll}
u_{0}(q, p, t) & c(q, p, t) \\
c^{\star}(q, p, t) & u_{1}(q, p, t)
\end{array}\right).
\label{eq:cqqubitBackground}
\end{equation} 
The dynamics of the CQ qubit has been studied in \cite{oppenheim2020objective}.  Moving away from states, we can define any {\it CQ operator} $f(z)$ which associates to each point in phase space a quantum operator.

It has been shown~\cite{oppenheim_post-quantum_2018} that any bounded dynamics mapping CQ states onto themselves, if taken to be linear, will be completely positive if and only if it can be written in the form
 \begin{equation}\label{eq: CPmap}
 \cqstate(z,t+ \tau) = \int dz'\sum_{\mu\nu} \Lambda^{\mu \nu}(z|z',\tau) \lin_{\mu} \cqstate(z',t) \lin_{\nu}^{\dag},
\end{equation}  
where for each $z,z'$ the $L_{\mu}$ are an orthogonal basis of operators (called \textit{Lindblad operators}) on the Hilbert space and $\Lambda^{\mu \nu}(z|z',\tau)$ is a positive, Hermitian matrix in $\mu \nu$ for each $z,z'$. More precisely $\int dz dz' A^*_{\mu}(z,z') \Lambda^{\mu \nu}(z|z')A_{\nu}(z,z') \geq 0$ for any vector $A_{\mu}(z,z')$. In equation \eqref{eq: CPmap}, the $\int dz$ integral is an integral over the phase space variables. For example, if $z=(q,p)$ then the integral is over $\int dq dp$. In the gravitational case we take $z=(g_{ab}, \pi^{cd})$ and the integral over $z$ will be a formal integral over all configurations of Riemmanian 3 metrics $g_{ab}$ and their conjugate momenta $\pi^{ab}$,  $\int \mathcal{D} g \mathcal{D} \pi$. We take this as a formal expression and we do not make any attempt to prove its existence. The normalization of probabilities requires
\begin{equation}\label{eq: probNorm}
\int dz  \sum_{\mu\nu}\Lambda^{\mu \nu}(z|z',\tau) \lin_{\nu}^{\dag}\lin_{\mu} =\mathbb{I}
\end{equation}
since the right hand side of Equation \eqref{eq: CPmap} must also be normalised. Equation \eqref{eq: CPmap}, combined with the normalization condition \eqref{eq: probNorm}, can be viewed as a generalisation of the classical transition probability equation%\footnote{The probability density of being at a point $z$ in phase space, after some time interval $\tau$ is given by $\rho(z,t+\tau)=\int dz' P(z|z';\tau)\rho(z',t)$ with $P(z|z';\tau)$ the transition probability.}
and the quantum Kraus decomposition theorem \cite{kraus2} to the hybrid case. For the unfamiliar reader, we compare the general dynamics for classical, quantum and classical-quantum dynamics, as well as their associated positivity and norm conditions, in Table \ref{tab: CPmapTable}. Henceforth, we will adopt the Einstein summation convention so that we can drop $\sum_{\mu\nu}$ with the understanding that equal upper and lower indices are presumed to be summed over.

\begin{table}
\begin{center}

\begingroup
\setlength{\tabcolsep}{10pt} % Default value: 6pt
\renewcommand{\arraystretch}{1.4}

\begin{tabular}{|c| c|} 
 \hline
 &  Classical \\  
 \hline 
Dynamics &  $p(z,t+ \tau) = \int dz' P(z|z',\tau) p(z,t)$  \\ 
 \cline{0-0}
Positivity condition &   $P(z|z', \tau) \geq 0 $ for all $z, z'$ \\ 
 \cline{0-0}
Norm condition &   $\int dz P(z|z', \tau) =1  $ \\ 
 \hline
  \hline
  &  Quantum \\
 \hline 
 Dynamics  &   $\sigma(t+\tau) = \sum_{\mu \nu} \lambda^{\mu   \nu}(\tau)L_{\mu} \sigma(t) L_{\nu}^{\dag} $  \\  
 \cline{0-0}
Positivity condition &   $\lambda^{\mu \nu}(\tau) $ a positive matrix in $\mu \nu$  \\ 
  \cline{0-0}
Norm condition &    $
\sum_{\mu \nu} \lambda^{\mu \nu}(\tau) L_{\nu}^{\dag} L_{\mu} =\mathbb{I} $ \\ 
 \hline
  \hline
  &   Classical-quantum \\
 \hline
 
Dynamics &  $\cqstate(z,t+\tau) = \int dz' \sum_{\mu \nu} \Lambda^{\mu \nu}(z|z',\tau)L_{\mu} \cqstate(z',t) L_{\nu}^{\dag} $ \\ 
 \cline{0-0}
Positivity condition &   $\Lambda^{\mu \nu}(z|z', \tau) $ a positive matrix in $\mu \nu$ for all $z, z'$ \\ 
 \cline{0-0}
Norm condition &  $
\sum_{\mu \nu} \Lambda^{\mu \nu}(z|z',\tau) L_{\nu}^{\dag} L_{\mu} =\mathbb{I} $ \\ 
 \hline
\end{tabular}
\caption{\label{tab: CPmapTable} A table illustrating the general dynamics governing finite time stochastic classical, quantum and classical-quantum dynamics.  
In the classical case, the probability $p(z)$ of being at a point $z$ in phase space, after some time interval $\tau$ is given in terms of the earlier probability distribution and  the transition probabilities $P(z|z';\tau)$ which define a classical channel.
We also show the positivity conditions which are required in order for dynamics to maintain positive probabilities, as-well as the norm condition which ensures probabilities sum to one. In this sense, one sees that equation \eqref{eq: CPmap} is a natural classical-quantum generalization of the classical transition probability equation and the quantum Kraus decomposition theorem to the hybrid case. }
\endgroup

\end{center}
\end{table}

%When the dynamics is Markovian o
One can derive the most general, linear, CQ master equation by performing a short time expansion of \eqref{eq: CPmap} \cite{blanchard1995event, diosi2014hybrid, alicki2003completely}.\footnote{ In the case where the classical degrees of freedom are taken to be discrete, Poulin~\cite{poulinKITP} was the first to use the diagonal form of Equation \eqref{eq: CPmap} to derive the master equation for bounded operators. The off-diagonal form is required in the continuous case\cite{oppenheim_post-quantum_2018,UCLPawula}.} To arrive at the master equation, we first introduce an arbitrary basis of traceless Lindblad operators on the Hilbert space, $L_{\mu} = \{L_0=I, L_{\alpha}\}$. At $\tau =0$ we know \eqref{eq: CPmap} is the identity map, which tells us that $\Lambda^{00} (z|z', \tau=0) = \delta(z,z')$. Looking at the short time expansion coefficients, by Taylor expanding in $\tau  \ll 1$, we can write
\begin{align}\label{eq: shortime}
&\Lambda^{\mu \nu}(z|z',\tau ) = \delta^{\mu}_{0} \delta^{ \nu}_{0} \delta(z,z') + W^{\mu \nu}(z|z') \tau + O(\tau^2).
\end{align}
By substituting the short time expansion coefficients into \eqref{eq: CPmap}, using the normalization condition \eqref{eq: probNorm}, and taking the limit $\tau \to 0$, we arrive at the classical-quantum master equation 
\begin{align}
\frac{\partial \cqstate(z,t)}{\partial t} =  \int dz' \ W^{\mu \nu}(z|z') L_{\mu} \cqstate(z') L_{\nu}^{\dag} - \frac{1}{2}W^{\mu \nu}(z) \{ L_{\nu}^{\dag} L_{\mu}, \cqstate \}_+,
\label{eq: cqdyngen}
\end{align}
where the bracket $\{, \}_+$ is quantum anti-commutator, which should be distinguished from the classical Poisson bracket which we write as $\{, \}$, and we have use the shorthand
\begin{equation}
W^{\mu  \nu}(z) = \int \mathrm{d} z' W^{\mu \nu}(z'| z).
\end{equation}
One can easily verify that \eqref{eq: cqdyngen} is trace preserving, and together with the  condition that
\begin{equation} \label{eq: block1}
\Lambda^{\mu \nu}(z|z',\tau) 
= \begin{bmatrix}
    \delta(z,z') + \tau W^{00}(z|z')      & \tau W^{0\beta}(z|z') \\
    \tau W^{\alpha 0}(z|z')    & \tau W^{\alpha \beta} (z|z')  \\
\end{bmatrix} + O(\tau^2)
\end{equation}
be a positive matrix in $\mu, \nu$ for all $z,z'$, this guarantees that the dynamics preserves probabilities. Moreover, it can be shown that equation \eqref{eq: cqdyngen}, combined with the positivity condition \eqref{eq: block1}, defines completely positive Markovian (memoryless) CQ dynamics for an arbitrary set of Lindblad operators $\{L_{\mu} \}$ \cite{oppenheim_post-quantum_2018}. The Markovian condition is natural if the dynamics are taken to be fundamental rather than effective, while in the more general case, the $W^{\alpha \beta}$ can be time-dependent, and need not be the elements of a  positive semi-definite matrix (see for example \cite{Breuer2012Foundations} in the case of the Lindblad equation). 

We see the CQ master equation \eqref{eq: cqdyngen} is a natural generalisation of the Lindblad equation and classical Pauli rate equation in the case of classical-quantum coupling. For the reader unfamiliar with the dynamics of open classical/quantum systems we compare the most general master equations for classical, quantum and CQ dynamics in Table \ref{tab: MasterEquationTable}.

Ultimately, we are interested in studying classical gravity coupled to quantum fields and so we shall need to consider the case in which the Lindblad operators are not bounded. In this case one can cast the master equation in the form of \eqref{eq: cqdyngen}, but it will be defined only on a subset of physical classical-quantum states for which the map \eqref{eq: cqdyngen} is a bounded map, so that its left hand side defines a normalized CQ state. For the purposes of the paper, we shall assume we work with such physical states. Furthermore, following standard convention, we refer to $W^{\mu \nu}(z|z')L_{\mu} \cqstate(z') L_{\nu}^{\dag}$ as the {\it jump term} and $ \frac{1}{2}W^{ \mu \nu}(z) \{ L_{\nu}^{\dag} L_{\mu}, \cqstate \}_+$ as the {\it no-event term}.\footnote{These conventions come from studying the unravellings of Lindblad equations via stochastic pure state quantum trajectories, where at each time step the quantum state either undergoes continuous evolution via an effective Hamiltonian, or with some probability jumps to a new state which on the Lindblad operators $L_{\alpha}$ appearing in the master equation \cite{Belavkin, Gardiner, Dalibard, breuerbook}. }

\begin{table}
\begin{center}
\begingroup
\setlength{\tabcolsep}{10pt} % Default value: 6pt
\renewcommand{\arraystretch}{1.4}
\resizebox{\columnwidth}{!}{
\begin{tabular}{|c| c|}
 \hline
 & Classical \\
 \hline
Master equation & $\frac{\partial p}{\partial t}=\int d z^{\prime} W\left(z | z^{\prime}\right) p\left(z^{\prime}\right)-\int d z^{\prime} W\left(z^{\prime}| z\right) p(z)$  \\ 
 \cline{0-0}
Positivity condition &    $P(z|z', \tau) = \delta(z,z') + \tau W(z|z') + \mathcal{O}(\tau^2) \geq 0  $ $\forall z,z'$ \\ 
 \hline
  \hline
  & Quantum \\
 \hline
Master equation  &  $\frac{\partial \sigma(t)}{\partial t}=-i[H, \sigma]+h^{\alpha \beta} L_{\alpha} \sigma L_{\beta}^{\dagger}-\frac{1}{2}\left\{h^{\alpha \beta} L_{\beta}^{\dagger} L_{\alpha}, \sigma\right\}_+$  \\ 
 \cline{0-0}
Positivity condition & $h^{\alpha \beta}$ a positive matrix, $h \succeq 0$.\\ 
 \hline
  \hline
  &   Classical-quantum \\
 \hline
Master equation & $\frac{\partial \varrho}{\partial t}=\int d z^{\prime} W^{\mu \nu}\left(z | z^{\prime}\right) L_{\mu} \cqstate\left(z^{\prime}\right) L_{\nu}^{\dagger}-\frac{1}{2}\int d z^{\prime} W^{\mu \nu}\left(z^{\prime} | z\right) \{L_{\nu}^{\dagger} L_{\mu}, \cqstate(z)\}_+$ \\
 \cline{0-0}
Positivity condition &   $\Lambda^{\mu \nu}\left(z | z^{\prime}, \tau \right)=\scalemath{0.8}{\left[\begin{array}{cc}\delta\left(z, z^{\prime}\right)+\tau W^{00}\left(z \mid z^{\prime}\right) &\tau W^{0 \beta}\left(z \mid z^{\prime}\right) \\ \tau W^{\alpha 0}\left(z \mid z^{\prime}\right) & \tau W^{\alpha \beta}\left(z \mid z^{\prime}\right)\end{array}\right]} + \mathcal{O}(\tau^2) $ $\succeq 0$ $\forall z,z'$ \\ 
 \hline
\end{tabular}}
\caption{\label{tab: MasterEquationTable} A table illustrating the master equations governing classical, quantum and classical-quantum dynamics. When the positivity conditions are satisfied, the master equations are Markovian (memoryless). We see that the CQ master equation is a natural generalization of the classical rate equation and the Lindblad equation. }
\endgroup
\end{center}
\end{table}

We can write the dynamics defined in \eqref{eq: cqdyngen} in an illuminating form by introducing a \textit{Kramers-Moyal expansion} of the master equation. In particular, defining
\begin{equation}\label{eq: definitionMoments}
D^{\mu \nu}_{n,i_1, \dots i_n}(z') = \frac{1}{n!} \int dz W^{\mu \nu}(z|z')(z-z')_{i_1} \dots (z-z')_{i_n} 
\end{equation}
the master equation in \eqref{eq: cqdyngen} can be written in the form
\begin{align}\label{eq: expansion}\nonumber
\frac{\partial \cqstate(z,t)}{\partial t} &= \sum_{n=1}^{\infty}(-1)^n \left(\frac{\partial^n }{\partial z_{i_1} \dots \partial z_{i_n} }\right) \left( D^{00}_{n, i_1 \dots i_n}(z) \cqstate(z,t) \right)\\ \nonumber
&  -i  [H(z),\cqstate(z)  ] + D_0^{\alpha \beta}(z) L_{\alpha} \cqstate(z) L_{\beta}^{\dag} - \frac{1}{2} D_0^{\alpha \beta} \{ L_{\beta}^{\dag} L_{\alpha}, \cqstate(z) \}_+ \\ 
& + \sum_{\mu \nu \neq 00} \sum_{n=1}^{\infty}(-1)^n  \left(\frac{\partial^n }{\partial z_{i_1} \dots \partial z_{i_n} }\right)\left( D^{\mu  \nu}_{n, i_1 \dots i_n}(z) \lin_{\mu} \cqstate(z,t) \lin_{\nu}^{\dag} \right),
\end{align}
where we define the Hermitian operator $H(z)= \frac{i}{2}(D^{\mu 0}_0 L_{\mu} - D^{0 \mu}_0 L_{\mu}^{\dag}) $ (which  is Hermitian since $D^{\mu 0}_0 = D^{0 \mu *}_0$, which follows from the Hermiticity of $\Lambda^{\mu \nu}(z|z')$). We see the first line of \eqref{eq: expansion} describes purely classical dynamics, the second line describes pure quantum Lindbladian evolution and is described by the zeroth moments; specifically the (block) off diagonals, $D^{\alpha 0}_0(z)$, describe the pure Hamiltonian evolution, whilst the components $D^{\alpha \beta}_0(z)$ describe the dissipative part of the pure quantum evolution. The third line contains the non-trivial classical-quantum interaction, where changes in the distribution over phase space are accompanied by changes in the quantum state. equation \eqref{eq: expansion} describes the most general form of CQ dynamics which is linear and Markovian. We shall now study a subset of such dynamics which we use when studying theories of gravity, namely dynamics which becomes Hamiltonian in the classical limit. 
\subsection{Hybrid dynamics with a Hamiltonian limit}\label{sec: CQham}
While the master equation \eqref{eq: cqdyngen} is completely general, we wish to restrict to dynamics which becomes approximately Hamiltonian in the classical limit. This was introduced in \cite{oppenheim_post-quantum_2018} and we shall review the formalism, since it is crucial in constructing CQ theories of gravity; in particular those which reproduce the Hamiltonian formulation of Einstein gravity once we also take the classical limit of the quantum system. We take the classical degrees of freedom to live in a phase space $\Gamma = (\mathcal{M}, \omega)$, where $\omega$ is the symplectic form. We further assume the pure classical evolution to be generated by a classical Hamiltonian $H_c$. While we believe one may need to consider dynamics where the purely classical dynamics is also stochastic, we here consider the restrictive and simpler case where it is deterministic. In this case we can write equation \eqref{eq: expansion} as
\begin{align}\label{eq: expansion2}\nonumber
\frac{\partial \cqstate(z,t)}{\partial t} &= \{ H_c, \cqstate(z,t) \}   -i  [H(z),\cqstate(z)  ] + D_0^{\alpha \beta}(z) L_{\alpha} \cqstate(z) L_{\beta}^{\dag} - \frac{1}{2} D_0^{\alpha \beta} \{ L_{\beta}^{\dag} L_{\alpha}, \cqstate(z) \}_+ \\ 
& + \sum_{\mu \nu \neq 00} \sum_{n=1}^{\infty}(-1)^n  \left(\frac{\partial^n }{\partial z_{i_1} \dots \partial z_{i_n} }\right)\left( D^{\mu  \nu}_{n, i_1 \dots i_n}(z) \lin_{\mu} \cqstate(z,t) \lin_{\nu}^{\dag} \right).
\end{align}

Now, we can define a phase space Hamiltonian vector field via  $X_h^{\alpha \beta, i} =  (\omega^{-1})^{ij} d_j h^{\alpha \beta }$. Here, $h^{\alpha \beta}(z)$ is some phase space functional and  $d_i$ is the exterior derivative on the phase space. Here, the form of $h^{\alpha \beta}(z)$ is motivated by wanting to reproduce an interacting Hamiltonian of the form 
\begin{align}
    H_I(z) = h^{\alpha \beta} L^{\dag}_{\beta} L_{\alpha}.
    \label{eq: ham decomposition}
\end{align} 
This choice will lead to a class of dynamics which we call the ``discrete class'' \cite{UCLPawula}, and it is the one which we will consider in the present work. The constraint algebra for the other class of dynamics, the ``continuous class" will be presented elsewhere \cite{oppenheim2021constraints}.
 
 By picking the interaction term $W^{\mu \nu}(z|z')$ in \eqref{eq: cqdyngen}to be such that the vector of first moments takes the form $D_{1i}^{\alpha \beta}(z) = X_h^{\alpha \beta,j}$, the interacting part of the dynamics in \eqref{eq: expansion2} becomes
\begin{equation}
\begin{split}
 & D^{\alpha \beta}_0(z) \lin_{\alpha} \cqstate(z, t)\lin_{\beta}^{\dag}  -\frac{1}{2} D_0^{\alpha \beta}(z)\left\{\lin_{\beta}^{\dagger} \lin_{\alpha}, \cqstate(z, t)\right\}_+ + \{ h^{\alpha \beta }, \lin_{\alpha} \cqstate \lin_{\beta}^{\dag}  \} \\
 & + \frac{\partial^2}{\partial z_{i_1} \partial z_{i_2}}(D^{\alpha \beta}_{2, i_1 i_2} \lin_{\alpha} \cqstate(z,t) \lin_{\beta}^{\dag}) + \dots \ .
\end{split}
\end{equation}
We see that defining $H_I(z)= h^{\alpha}(z) \lin_{\alpha}^{\dag} \lin_{\alpha}$ and taking the trace over the quantum system gives us an effective classical equation of motion
\begin{equation}
\frac{\partial p(z,t) }{\partial t} = \{H_c, \cqstate(z,t) \} + \Tr{}{\{H_I(z) , \cqstate(z) \}} + \dots \ .
\label{eq: limit}
\end{equation}
Where the $\dots$ denote the higher order terms in the moment expansion. In order to reproduce Hamiltonian dynamics we imagine the scenario in which the higher order moments in \eqref{eq: limit} are suppressed by some order parameter \cite{oppenheim_post-quantum_2018, oppenheim2020objective}. For clarity, we can compare this limit to the case where we have two classical systems, $(z_1, z_2)$, which interact via an interaction Hamiltonian $H_I$. If the dynamics of the total system is given by 
\begin{equation}
\frac{\partial p(z_1,z_2,t) }{\partial t} = \{ H_1, \rho \} + \{ H_2, \rho \} + \{ H_I, \rho \}
\end{equation} 
then integrating out the second system, and defining $\bar{\rho}(z_1) = \int dz_2 \rho(z_1,z_2)$, we get an effective equation of motion 
\begin{equation}
\frac{\partial \bar{\rho} (z_1)}{\partial t} = \{ H_1, \bar{\rho} (z_1) \} + \int \mathrm{d} z_2 \{ H_I , \rho(z_1,z_2)\},
\end{equation}
which justifies \eqref{eq: limit} as the appropriate classical limit to ask for in order to reproduce Hamiltonian dynamics. It also ensures the classical degrees of freedom undergo Hamiltonian evolution on average -- a type of Eherenfest theorem for CQ dynamics.  There are of course some ambiguities in the construction, for example there is no unique decomposition of the Hamiltonian into Lindblad operators $L_{\alpha}$, but this shall not be relevant for the discussion at hand. 
A natural question, then, is to ask whether or not this formalism can be used to construct a consistent theory of CQ gravity, where the gravitational field is taken to be classical, whilst the matter fields are considered to be quantum. In particular, the CQ theory should reproduce general relativity in the classical limit of the quantum system. The subtleties arise since gravity is a gauge theory, not only do the degrees of freedom undergo dynamics generated by the pure gravity Hamiltonian but they must also live on the constraint surface. Hence, although we can use the formalism introduced in this section to construct CQ dynamics which reproduces that of gravity, we must find a way of studying the constraints in CQ theories.

\section{CQ theory which reproduces Einstein gravity}\label{sec: CQeinstein}
In this section, we review how one can use the CQ formalism introduced in the previous section to construct CQ models of gravity which reproduce Einstein gravity \cite{oppenheim_post-quantum_2018}. We here restrict ourselves to the case where the CQ equation has the purely classical evolution generated by the ADM Hamiltonian, the pure quantum evolution generated by the Klein-Gordon (KG) Hamiltonian and the interaction term a CQ dynamics, whose first moment is such that it approximates the Hamiltonian formulation of gravity in the classical limit. 
\subsection{Liouville formulation of classical gravity}\label{sec: louiville}
In pure classical gravity, the relevant configuration space is given by the space of Riemannian metrics on a surface $\Sigma$, which we denote $Q$. The phase space is then given by $TQ^*$, endowed with its canonical symplectic form, derived from the exterior derivative of the tautological one form on $TQ^*$ \cite{ashtekar1982}. Elements of $TQ^*$ are denoted $(g_{ab}, \pi^{ab})$ and are taken to satisfy the canonical Poisson bracket relations\footnote{The convention here is that $\delta(x,y)$ is a scalar in $x$ and a scalar density in $y$. It is defined by its action on scalar functions $f : \Sigma \to \mathbb{R}$ as
$
f(x) = \int_{\Sigma} dy \delta(x,y) f(y)
$. It is useful to note that as a consequence, $\int_{\Sigma} dy \nabla_a^x \delta(x,y) f(y) = \nabla_a f(x)$ and $\int_{\Sigma} dy \nabla_a^y \delta(x,y) f(y)= -\int_{\Sigma} dy  \delta(x,y) \nabla_a^x f(y) = - \nabla_a f(x) $.}
\begin{align}
\{ g_{ab}(x), \pi^{cd}(y) \} = \frac{1}{2}( \delta_{a}^{c} \delta_b^d + \delta_a^d \delta_b^c) \delta(x,y).
\end{align}
The dynamics is generated by the ADM Hamiltonian \cite{arnowitt2008republication, dewitt1967quantum}
\begin{align}
H_{ADM}[N,\vec{N}]=\int d^3x (N\lapse+N^a\cm_a) = H[N] + H[\vec{N}]
\label{ADM}
\end{align}
where
\begin{align}
\lapse=
%\frac
{(16\pi G)}
%{c^6}
\pi^{ab} G_{abcd}\pi^{cd}-\frac{1}{16\pi G}\g^{1/2}R, \s \cm_a=-2\g_{ac} D_b\pi^{cb},
\label{eq:superham}
\end{align}
 $G_{abcd}$ is the deWitt metric defined as $G_{abcd}=\frac{1}{2\sqrt{g}}(g_{ac}g_{bd}+g_{ad}g_{bc}-g_{ab}g_{cd})$ and $D_a$ the covariant derivative with respect to the metric $g_{ab}$ on $\Sigma$.\footnote{ We extend the covariant derivative to act on tensor densities of weight $W$, we do this by subtracting $W \Gamma_{c}^{ca}$ to the usual covariant derivative $D_a$ For example $D_a \pi^{bc} = \partial_a \pi^{bc} + \Gamma^b_{ad} \pi^{dc} + \Gamma^c_{ad} \pi^{bc} - \Gamma^d_{da} \pi^{bc}$ since $\pi^{ab}$ is a tensor density of weight 1.}

The lapse function $N$ and shift vector $N^a$ appearing in $\eqref{ADM}$ are arbitrary functions of $(t,x)$.  They arise when performing the $3+1$ split of space-time and represent the gauge degrees of freedom associated to picking a foliation of space-time. They are seen to be non-dynamical, since $P_N, P_{N^a}=0$, and as a result the Hamiltonian formulation of GR is a constrained theory. Asking that the constraints $P_N, P_{N^a}=0$ be preserved in time leads to the Hamiltonian and Momentum constraints, $\mathcal{H} = \cm_a =0$. Conservation of these constraints is ensured via the hypersurface deformation algebra
\begin{equation}
\begin{array}{l}

\{H[N], H[M])\}=H[\vec{R}]\\
\{ H[\vec{M}], H[N] \}=H\left[ L_{\vec{M}} N \right]\\
\{ H[\vec{N}], H[\vec{M}] \} = H[ L_{\vec{N}} \vec{M}],
\end{array}
\label{eq: defalg}
\end{equation}
where $R^{a}:=g^{a b}\left(N D_b M -M D_b N\right)$ and  $L$ is the Lie derivative on $\Sigma$.

Since the language of CQ dynamics is that of master equations, it is worth mentioning that -- although it is not usually considered -- we can write the dynamics of pure GR in a Liouville formulation. In particular, a phase space distribution $\rho(g,\pi)$ will evolve under the dynamics as 
\begin{equation}
\frac{\partial \rho }{\partial t} = \{H_{ADM}, \rho \},
\end{equation}
subject to the constraints $\mathcal{H} \rho = \cm_a \rho =0$; that is $\rho$ must have support only on the constraint surface. There are a few things to be wary of when using a Liouville  formalism. Firstly, we must remember that $\rho$ is a distribution and so its action is only defined once smeared over phase space test functions. Furthermore, in the Liouville picture we solve for $\rho( g, \pi,t)$ and the solution can be interpreted as describing a probability density over trajectories $(g_{ab}(t,x), \pi^{cd}(t,x))$, which, using the ADM formalism, we can use to define a probability distribution over 4-geometries $g_{\mu \nu}(t,x)$. Since the trajectories are deterministic, and we can imagine starting in a state of certainty $\rho \sim \delta(\bar{g},g)\delta(\bar{\pi},\pi)$, we should then be able to pick a different lapse and shift for each point in phase space $N(t, g, \pi), N^a(t, g, \pi)$. A question then arises: should $N(t, g, \pi), N^a(t,g, \pi)$ be taken to be in or outside the Poisson bracket? 
In GR it turns out that this distinction does not matter, since if we were to include $N, N^a$ inside the Poisson bracket we have 
\begin{equation}
\frac{\partial \rho }{\partial t} =\int d^3x  \{ N(g,\pi), \rho \} \mathcal{H} + N(g,\pi) \{ \mathcal{H}, \rho \} + \{ N^a(g,\pi), \rho \} \mathcal{H}_a + N^a(g, \pi)\{ \mathcal{H}_a, \rho \} ,
\label{eq: lapseshift}
\end{equation}
so the difference between the equations of motion in \eqref{eq: lapseshift}, and the equations of motion if $N, N^a$ were not to be included in the Poisson bracket, vanishes on the constraint surface. In the CQ theory, however, this will not be guaranteed and so the distinction becomes potentially important.  If the lapse has an explicit dependence on the metric, or its conjugate momenta, then conservation of probability dictates it needs to appear \textit{inside} the Poisson bracket, and any higher derivatives of equation \eqref{eq: expansion}. Here, we take the lapse and shift to only depend on $x,t$ and discuss the more general case in appendix \ref{sec: realisations}.

We can of course add matter to the discussion. For ease of calculation, we shall here be interested in coupling scalar fields to gravity. A classical field minimally coupled to gravity  will have a Hamiltonian of the form
\begin{equation}
H_T\gauge = H_{ADM}\gauge + H_{m}\gauge= \int d^3 x N( \ch + \ch_m) + N^a( \cm_a + \cm_{m,a}),
\label{GRtot}
\end{equation}
where $H_m$ is the Hamiltonian of the matter field. In the presence of matter, the constraint surface takes the form $\ch + \ch_m =0, \cm + \cm_{m,a} =0$. 

For example, the Hamiltonian of the free scalar field reads
\begin{equation}
H_m[N,\vec{N}]=\int d^{3} x N(\frac{1}{2}g^{-1/2} \pi^{2} +\frac{1}{2} g^{1/2} g^{i j} \partial_{i} \phi \partial_{j} \phi+\frac{1}{2}g^{1/2} m^{2} \phi_{\phi}^{2})+N^{i}\pi_{\phi} \partial_{i} \phi.
\label{eq: scalar field ham}
\end{equation}
The Liouville equation for the phase space density $\rho(g, \pi_g , \phi, \pi_{\phi},t)$ then takes the form
\begin{equation}
\frac{\partial \rho }{\partial t} = \{H\gauge, \rho \} + \{ H_m\gauge, \rho \},
\label{GR}
\end{equation}
where $\rho$ must only have support on the constraint surface. 

In order to gain parallels to the CQ theory, in particular when looking at the gravitational degrees of freedom alone, it is insightful to integrate out the $\phi, \pi_{\phi}$ degrees of freedom to get an effective equation for the evolution of the gravitational degrees of freedom. Defining $ \bar{\rho}( g, \pi_g) = \int D \phi D_{\pi_{\phi}} \rho(g, \pi_g , \phi, \pi_{\phi})$ then integrating \eqref{GR} over the matter degrees of freedom gives an effective equation of motion 
\begin{equation}
\frac{\partial \bar{\rho}( g, \pi_g,t) }{\partial t} = \{H, \bar{\rho} \} + \int D \phi D_{\pi_{\phi}} \{ H_m, \rho \} ,
\label{effective}
\end{equation}
which is the version of \eqref{eq: limit} if matter were to be treated classically. 

\subsection{CQ theories of gravity}\label{sec: cqtheoriesintro}

 In line with the discussion of section \ref{sec: CQham}, we are able to construct CQ theories of gravity whose dynamics becomes approximately that of Einstein gravity in the classical limit. To be slightly more precise: following \cite{oppenheim_post-quantum_2018} and the toy models in \cite{oppenheim2020objective}, the interaction between the classical and quantum degrees of freedom causes the quantum state to change, while at the same time causing back-reaction on the classical degrees of in a way which approximates Hamiltonian evolution generated by the ADM Hamiltonian. To that end, we consider the CQ equation in \eqref{eq: expansion2}, where we take the pure classical dynamics to be generated by $H_{ADM}$, whilst we take the pure quantum evolution to be generated by the Klein-Gordon Hamiltonian $H_m$
\begin{equation}
H_m\gauge = \int d^3 x N h^{\ag \bg} \lin_{\bg}^{\dag} \lin_{\ag} + N^a p_a^{\ag \bg} \lin_{\bg}^{\dag} \lin_{\ag} = \int d^3 x N(x) \mathcal{H}_m(x) + N^a(x) \qm_{m,a}(x),
\label{matterham}
\end{equation}
which we now take to be a quantum object. Here we have defined
\begin{equation}
\begin{gathered}
 h^{\pi \pi}:= \frac{1}{2} g^{-1/2}, \quad h^{\phi \phi} = \frac{1}{2} g^{1/2}, \quad h^{ab} = \frac{1}{2}g^{1/2} g^{ab} 
 \label{eq: h for scalar field}\\ 
p^{a \pi} = 1/2, \quad p^{\pi a} = 1/2 \\
 \lin_{\pi}(x)= \pi_{\phi}(x), \quad \lin_{\phi}(x)= \phi(x), \quad \lin_{a}(x)=\partial_{a} \phi(x),
\end{gathered}
\end{equation}
and the, now quantum, field operators satisfy the canonical commutation relations
\begin{align}\label{eq: commutation}
[ \phi(x), \phi(y)] = 0, \s[ \pi_{\phi}(x), \pi_{\phi}(y)] = 0, \s [ \phi(x), \pi_\phi(y)]=i \delta(x,y).
\end{align} 
Again, the $\delta(x,y)$ is defined such that it is a scalar in $x$ and a scalar density in $y$. We shall not construct the precise nature of the Hilbert space $\mathcal{H}$, ultimately in the calculations we only exploit the algebraic properties of the canonical commutation relations and so the details are not of primary importance here. However, let us briefly comment on the problem of defining the Hilbert space 

Even for a free field in a fixed background $g_{\mu \nu}$ there exists infinitely many unitarily in-equivalent representations of the commutation relations on a Hilbert space, and in general there is no notion of a preferred state for which a Hilbert space representation can be defined \cite{Haag1993LocalQP, Hollands:2014eia}. The modern view is to take an algebraic approach to quantum fields in curved space. One instead views the algebraic (commutation) relations satisfied by the field observables as fundamental, which are taken to belong to an algebra $\mathcal{A}$. States $\omega$ are then defined as positive linear functionals on $\mathcal{A}$. In this view point, the algebraic structure is unique, but there are infinitely many \textit{representations} of the algebra on a Hilbert space. In particular, the GNS construction \cite{Gelfand, Segal} shows that every state $\omega$ on the algebra defines a Hilbert space $\mathcal{H}$, a representation of the algebra on the Hilbert space, and a Hilbert space vector corresponding to $\omega$. Thus, though technically equivalent, the algebraic approach allows one to formulate QFT in curved space in a way which is independent of the representation of the algebra, and does not require one to single out a preferred state in the theory.

In the classical-quantum case, there is a similar problem in defining the Hilbert space. It is simple enough for us to define some preferred Hilbert space $\mathcal{H}$ and to take the field operators to act on $\mathcal{H}$. As an example, we could take the standard Hilbert space of free field in Minkowski space, defined by the existence of a Poincare invariant vacuum state \cite{Hollands:2014eia}. We could then consider classical-quantum dynamics on this Hilbert space. However, this arbitrary choice of Hilbert space is clearly not satisfactory. Nonetheless, it's important to emphasize that we can view classical-quantum dynamics, and the commutation relations \eqref{eq: commutation}, in an algebraic way -- for example by using the Heisenberg representation of classical-quantum dynamics found in \cite{oppenheim_post-quantum_2018}.

Within this formulation, the problem of formulating an algebraic version of CQ dynamics reduces to finding the appropriate CQ version of axioms on the fields $\phi(x)$ so that they obey the CQ Heisenberg equations of motion and form a $C^*$ algebra, in which case the GNS construction can be used to define an appropriate Hilbert space. Since this is beyond the scope of this work, for now we view the operators acting on the Hilbert space as formal expressions, and we leave an algebraic formulation of classical-quantum dynamics as an interesting open problem for future research.

Now we have specified the pure classical and quantum dynamics, all that remains is to specify the classical-quantum interaction term which describes the back reaction of the quantum system on the classical degrees of freedom. We call a specific choice of the CQ coupling a \textit{realization} of a CQ theory. Firstly, we demand that the first moments of the Kramers-Moyal expansion in \eqref{eq: expansion2} contain a term 
\begin{equation}
\int d^3 x N(x) \frac{\delta h^{\ag \bg}}{\delta g_{cd}} L_{\alpha} \cqstate L_{\beta}^{\dag} 
\label{eq: firstmom}
\end{equation}
in order to yield Einstein gravity in the classical limit. As a consequence, we view \eqref{eq: firstmom} as a condition on CQ theories of gravity that a sensible realization of CQ dynamics must satisfy.  Here the $\alpha, \beta$ indices run over $\phi, \pi_{\phi}, a$. In order to study CQ dynamics in a concrete setting we shall make some further assumptions about the realizations. Firstly, we assume that the realizations are local, so that the CQ interaction is fully specified by the set of transition amplitudes $W^{\ag \bg}(z|z',x)$, $W^{\ag \bg}_a(z|z',x) $. Secondly, we shall focus on a natural class of dynamics, namely those with CQ couplings $W^{\ag \bg}(z|z',x)$ which are linear in $N$ and $N^a$. There is a good reason to do this: from a physical stand point, the free functions $N$ and $N^a$ represent the local time reperameterization invariance and space-like diffeomorphism invariance of the underlying theory. A natural question, is to ask whether or not we can have a CQ theory of gravity which upholds these symmetries. Furthermore, from a technical point of view if we were to consider non-linear couplings in the lapse and shift -- say we had an $N^2$ coupling -- then the method we use to derive the constraints would lead to a constraint which itself depends on $N$. Preservation of such a constraint then leads to $N$ becoming dynamical, which generically leads to a gauged fixed theory; this happens for example in Horava gravity \cite{horava, horavaphase,horavaphase2} (although there are exceptions, where additional secondary constraints fix $N$ up to a global reperameterization invariance, for example in shape dynamics \cite{mercati2017shape, Anderson_2005, Gomes_2012}). We leave gauge fixed CQ theories as a possible area for further study. 
%Making the assumption that realizations are linear in the lapse and shift, we are lead to study dynamics of the form
%\begin{equation}
%\begin{split}
%\frac{\partial\cqstate(g,\pi)}{\partial t}&=\int d^3xd^3y dz' W^{00}(z|z;x-y)
%\cqstate(z')
%-i [ H_m, \cqstate]  \\
%&+\int d^3x dz' \Big[(N W^{ \ag \bg} (z|z',x-y)+ N^a W_a^{\ag \bg}(z|z',x-y))\lin_{\ag}(x) %\cqstate(z') \lin_{\bg}^{\dag}(y) \big] \\
%&-\frac{1}{2 } \int d^3 xd^3 y[ (N W_0^{\ag \bg}(z;x-y) + N^a W_{0a}(z;x-y)^{\ag \bg} ) %\{\lin_\bg^\dagger(x))\lin_\ag(y),\cqstate\}
%\end{split}
%\label{eq:dynamicalreal_gen}
%\end{equation}

%where $\int d^3xd^3y dz' W^{00}(z|z;x-y)
%\cqstate(z')= \{ H_{ADM}, \cqstate\} +\cdots$ includes the deterministic pure gravity evolution but could also contain classical diffusion and other higher order terms. We also allow the couplings to include a regulator $\epsilon(x-y)$ as in \cite{bps,OR-intrinsic}. When  $\epsilon(x-y)$ only has support close to $0$, then the $W(z|z';x-y)$ become local functions of $x$, $W(z|z';x)$, while if $\epsilon(x-y)$ is broad, one has violations of cluster-decomposition\cite{bps,OR-intrinsic}. For the purposes of the present analysis, we will make the assumption that $\epsilon(x-y)$ is local, and we will further ignore the pure gravity stochastic terms so that $\int d^3x dz' W^{00}(z|z;x) = \{ H_{ADM}, \cqstate\}$. The more general case will be analysed elsewhere.
In line with section \ref{sec: CQham}, we will thus consider the simplified dynamics
\begin{equation}
\begin{split}
\frac{\partial\cqstate(z)}{\partial t}&= \{ H_{ADM}, \cqstate(z)\} -i [ H_m, \cqstate(z)] \\
& +\int dz' \int d^3x  \Big[(N W^{ \ag \bg} (z|z',x)+ N^a W_a^{\ag \bg}(z|z',x))\lin_{\ag}(x) \cqstate(z') \lin_{\bg}^{\dag}(x) \big] \\
&-\frac{1}{2 } \int d^3 x[ (N W_0^{\ag \bg}(z) + N^a W_{0a}(z)^{\ag \bg} ) \{\lin_\bg^\dagger(x))\lin_\ag(x),\cqstate(z)\}, 
\label{eq:dynamicalreal}
\end{split}
\end{equation}
where the first moments of the realizations satisfy equation \eqref{eq: firstmom}. Here, $z$ labels points in the phase space of GR, $z=(g_{ab}, \pi^{cd})$, and writing out equation \ref{eq:dynamicalreal} in full we obtain 
\begin{equation}
    \begin{split}
& \frac{\partial\cqstate(g, \pi)}{\partial t}= \{ H_{ADM}, \cqstate(g, \pi)\} -i [ H_m, \cqstate(g,\pi)] \\
& +\int \mathcal{D}g' \mathcal{D} \pi' \int d^3x  \Big[(N W^{ \ag \bg} (g,\pi|g', \pi',x)+ N^a W_a^{\ag \bg}(g, \pi|g',\pi',x))\lin_{\ag}(x) \cqstate(g',\pi') \lin_{\bg}^{\dag}(x) \big] \\
&-\frac{1}{2 } \int d^3 x[ (N W_0^{\ag \bg}(g, \pi) + N^a W_{0a}(g, \pi)^{\ag \bg} ) \{\lin_\bg^\dagger(x))\lin_\ag(x),\cqstate(g, \pi)\}.
\label{eq:dynamicalreal2}
\end{split}
\end{equation}
  In equation \eqref{eq:dynamicalreal2} the integral over $\int \mathcal{D} g \mathcal{D} \pi$ is to be treated as a formal integral over all configurations of Riemmanian 3 metrics $g_{ab}$ and their conjugate momenta $\pi^{ab}$, and we do not make any attempt to rigorously justify its existence. To simplify notation, we will often suppress the phase space integrals and write, for example, $ \int d z' W^{\ag \bg} ( z | z',x) \cqstate(z') = W^{\ag \bg}(x)(\cqstate) $ indicating that $W^{\ag \bg}(x)$ is a differential (or in other cases, a CQ CP map) acting on $\cqstate$. Furthermore, we will often write the master equation in a more compact form as 
\begin{align}
\frac{\partial\cqstate(g,\pi)}{\partial t}=
\int d^3x {N \mathcal{L}(\cqstate)} + { N^a \mathcal{L}_a (\cqstate) },
\label{eq:dynamicalPQG0}
\end{align}
implicitly defining
\begin{align}
& \begin{split}
\mathcal{L}(x)(\cqstate)  &= \{ \ch (x) , \cqstate \} -i[ \qh(x), \cqstate] \\
& + W^{ \ag \bg}(x)\lin_\ag(x)\cqstate\lin_\bg^\dagger(x) - \frac{1}{2 } W_0^{\ag \bg}(z,x)\{\lin_\bg^\dagger(x)\lin_\ag(x),\cqstate\}
\end{split}\\
& \begin{split}
\mathcal{L}_a(x)(\cqstate) &= \{ \cm_a(x), \cqstate \} -i[ \qm_{m,a}(x), \cqstate] \\
& + W_a^{\ag \bg}(x)\lin_{\ag}(x) \cqstate \lin_{\bg}^{\dag}(x) - \frac{1}{2 } W_{0 a}^{\ag \bg}(z,x)\{\lin_\bg^\dagger(x)\lin_\ag(x),\cqstate\}
\label{eq:gener0}.
\end{split}
\end{align}
Equation \eqref{eq:dynamicalreal} gives us a very natural class of hybrid theories which give the dynamics of GR in there classical limit. Of course things are actually more complicated than this, GR is a constrained system, not only do the degrees of freedom undergo dynamics generated by the ADM Hamiltonian, but they must also lie on the constraint surface. Hence, we expect that constraints must enter into any CQ theory of gravity. In the usual picture, the constraints come directly from an action principle -- which we do not have here. How then can we impose constraints on the CQ theory? It is known in classical GR it is possible to derive the constraints by exploiting the algebroid nature of the hypersurface deformation algebra \cite{hojman1976geometrodynamics} (see also \cite{diracbook,kucharkarel, leesmolincos, 2004boj, Thiemann:1996aw, Gaul:2000ba, Isham:1992ex, thiemann}, for discussion of the deformation algebra in other contexts). As it turns out, similar consistency conditions can be found by considering the Dirac argument -- if the equations of motion contain arbitrary functions of time, such as the lapse and shift, then in order to retain predictability, two solutions related by a different choice of the arbitrary function must be gauge equivalent \cite{diracbook}. Although the argument, in its original form, is applied to Hamiltonian dynamics arising from action principles, it naturally extends to the case of CQ theories. We outline this argument in the next section and show how it leads to a constraint surface. We then apply this to the CQ theory introduced in this section and use it to derive generalised Hamiltonian and momentum constraints in CQ theories of gravity. 
\section{Deriving constraints from gauge conditions}\label{sec: derivingconstraints}
In this section, we briefly review the Dirac argument \cite{diracbook} for gauge theories. We then show, by applying a Dirac like argument to GR, that one can arrive at a set of consistency conditions which lead to constraints on the theory. The mathematics is similar to that used in HKT \cite{hojman1976geometrodynamics}, which exploits the algebroid nature of GR to arrive at the constraint surface. We then extend this method to the case of CQ master equations. In particular, this provides a methodology to derive a set of constraints for CQ theories which are linear in $N, N^a$ which is a central result of this paper. 
\subsection{Dirac argument for Hamiltonian systems}\label{sec: dirac}
Hamiltonian gauge theories generically have actions of the form \cite{diracbook, banados}
\begin{align}
I[q^i, p_j, \lambda^a] = \int dt(p_i \dot{q}^i - H_0(p_i, q^j) - \lambda^a \phi_a( p_i, q_i)),
\end{align}
which define equations of motion and constraints
\begin{align}
\dot{q}^i = \frac{\partial H_0 }{\partial p_i} - \lambda^a \frac{\partial \phi_a}{\partial p_i}, \s \dot{p}_i = -\frac{\partial H_0 }{\partial q_i} + \lambda^a \frac{\partial \phi_a}{\partial q_i}, \s \phi_a =0.
\end{align}

Preservation of the constraints requires that
\begin{align}
\frac{d \phi_a}{dt } = [ \phi_a, H_0] - [ \phi_a, \phi_b]\lambda^b \approx 0.
\end{align}
%Here $\approx$ means that the equation must be weakly zero, i.e, it must vanish on the constraint surface. 
Gauge theories are characterized by having $[ \phi_a, \phi_b]  \approx 0$, in which case the constraints are said to be first class. In this case, the $\lambda^a(t)$ remain undetermined and the equations of motion contain arbitrary functions of time, as will there solutions. Gauge theories then generically have Hamiltonian's of the form
\begin{equation}
H_T = H_0 + \lambda^a(t) \phi_a,
\end{equation} 
where the $\lambda^a(t)$ are arbitrary and they multiply constraints. Usually, it is said that first class constraints generate equal time gauge transformations. To see why this is we can run the Dirac argument \cite{diracbook} as follows. Suppose we have some initial data $(q^i, p_j)$ at $t=0$. Since the $\lambda(t)$ are undetermined we can use either $\lambda(t)$ or $ \lambda'(t)$ to solve the equations of motion. Let $f(q^i, p_j)$ be any functional over the phase space. Then at time $t= \epsilon$ 
\begin{align}
&f( \epsilon) = f(0) + \epsilon \{f(0), H_0 \} + \epsilon \lambda^a \{ f(0), \phi_a \} \\
& f'( \epsilon) = f(0) + \epsilon \{f(0), H_0 \} + \epsilon \lambda^{a \prime} \{ f(0), \phi_a \} .
\end{align}
As a consequence, given the same set of initial data we get two equally valid solutions. If we wish to retain any notion of predictability we must identify these solutions as being physically equivalent. The difference between the two solutions is 
\begin{align}\label{eq: difference}
\delta f(\epsilon) = \epsilon( \lambda'- \lambda) \{ f(0), \phi_a \}
\end{align}
and hence $ \{ \phi_a \}$ generates equal time gauge transformations. More generally, whole solutions which are related by different choices of Lagrange multiplier should be taken to be gauge equivalent, which reduces to equation \eqref{eq: difference} when one looks at equal time slices of the solutions \cite{pons}.

Given that constraints generate gauge transformations, it should then be true that the generators form an algebra (which follows from linearity and demanding that the gauge transformations be transitive). For first class constraints we have 
\begin{align}
\{ \phi_a, \{ \phi_b, f \} \} - \{ \phi_b, \{ \phi_a, f, \} \} = \{C^c_{ab} \phi_c, f \} = \{ C^{c}_{ab}, f \} \phi_c + C_{ab}^c \{ \phi_c, f \} \approx  C_{ab}^c \{ \phi_c, f \},
\end{align}
so that the gauge generators close as an algebra on the constraint surface. Note this holds even for the case of GR, where the $C^{a}_{bc}$ are structure functions which depend on the phase space.

\subsection{Deriving the constraint surface of GR from the Dirac argument}\label{sec: constraintdirac}

 We now show how use of the Dirac argument for classical gravity leads to constraints. Suppose we have a state $\rho(g, \pi)$ which evolves according to the ADM equations of motion
\begin{align}
\frac{\partial \rho}{\partial t } = \int d^3x \{ N \mathcal{H} + N^a \mathcal{H}_a , \rho \}.
\end{align}
 For now, we will not take $N(t,x), N^a(t,x)$ to have a functional dependence on $g, \pi$, which is the weakest condition we can ask for. We will have to be careful since the states $\rho(g, \pi)$ are defined only in the distributional sense so things like Poisson brackets are only defined by there action on test functions. We denote the smeared distributions as
\begin{align}
\langle A, \rho\rangle = \int Dg D\pi \rho(g,\pi, t) A(g,\pi,t).
\end{align}
Since the lapse and shift functions are arbitrary $\{ \mathcal{H},\} $ and $\{ \mathcal{H}_a, \}$ generate equal time gauge transformations. As a consequence
\begin{align}
\langle A, \rho\rangle \sim \langle A, \rho\rangle  + \epsilon \langle  \{ A, \mathcal{H}_a(x) \}, \rho \rangle.
\end{align}
 Since all observables, $O(g,\pi,t)$ must be independent of the choice of gauge, they must satisfy $\langle  \{ O(g,\pi,t), \mathcal{H}_a(x) \}, \rho \rangle =0$. In particular, picking $\rho = \delta(\bar{g}, g) \delta( \bar{\pi},\pi)$, we see observables must satisfy
 \begin{equation}
\{ O, \mathcal{H}_a(x) \} \approx 0.
 \end{equation} Note, we are not saying that time evolution is a gauge transformation and observables are frozen. The gauge generators take the state of a solution to the equations of motion at time $t$ to another solution at the same time. Furthermore, allowing for the case where the observables contain an explicit time dependence -- which is what usually occurs once one has gauge fixed \cite{Pons_2010} -- means they evolve in time even though there Poisson bracket with the Hamiltonian vanishes. 
 
 Demanding that the gauge transformations close as an algebra, which follows from asking that the equivalence relation be linear and transitive, we must have
\begin{align}
\{ H[N], \{ H[M], \rho \} \} - \{ H[M], \{  H[N], \rho \} \} \approx \text{Gauge} (\rho).
\label{eq: compat}
\end{align}
When equation \eqref{eq: compat} is smeared over observables we see that the left hand side must vanish since we know that observables Poisson commute with the constraints.

We now use the Jacobi identity and the hypersurface deformation algebra defined in equation \eqref{eq: defalg} to write equation \eqref{eq: compat} as
\begin{align}
\{ \{ H[N], H[M] \} , \rho \} = \int d^3x \{  g^{ab} \delta N_b \mathcal{H}_a (x), \rho \} \approx \text{Gauge} (\rho),
\label{eq: classcons}
\end{align}
where we have defined $ \delta N_b = N \partial_b M - M \partial_b N$. Smearing equation \eqref{eq: classcons} over a phase space smearing function $A$, one finds
\begin{align}
\int Dg D\pi \int d^3x \{  g^{ab} \delta N_b \mathcal{H}_a (x), \rho \} A = \int d^3 x  \langle\delta N_b \{ A, g^{ab}\}, \mathcal{H}_a \rho \rangle + \langle \{A, \mathcal{H}_a \} \delta N_b g^{ab}, \rho \rangle .
\label{gr con}
\end{align}
The final term in equation \eqref{gr con} is a gauge transformation and so we see that for the algebra of gauge transformations to close 
\begin{equation}
    \langle \{A, g^{ab} \}, \mathcal{H}_a \rho \rangle
     \approx 0.
     \label{eq: classcon}
\end{equation} 
In other words, we require $\langle \{A, g^{ab} \}, \mathcal{H}_a \rho \rangle =0$ whenever $A$ is a phase space observable.
Clearly this condition is satisfied by the constraints of GR since $\langle A, \mathcal{H}_a  \rho \rangle=0$ for all phase space functions $A$ when on the constraint surface. Preservation of this condition is guaranteed by the Hamiltonian constraint  $\langle A, \mathcal{H}  \rho \rangle=0$, as can be seen by applying the evolution equation on it and using the Jacobi identity. However, the consistency condition \eqref{eq: classcon} is weaker than asking that the constraints hold. As a trivial example of this, we can consider a theory in which the only allowed observable is $A= I$, then \eqref{eq: classcon} holds for any $\rho$ on the phase space. In general, we shall ask the question: is there a set of sensible observables and states for which consistency conditions -- such as that in equation \eqref{eq: classcon} -- hold and which are preserved in time? This is similar to the \textit{off-shell closure} of Theimman \cite{Thiemann:1996aw}. For dynamics generated by the ADM Hamiltonian, it seems as though there is no non-trivial set of states/observables other than that of GR for which \eqref{eq: classcon} holds. The problem is that we can keep applying gauge transformations to generate more and more independent constraints. For example, suppose that \eqref{eq: classcon} is satisfied, then applying a spatial gauge transformation we also require that
\begin{equation}
\langle \{  \{ g^{ab}, A \}, \mathcal{H}_c \}, \mathcal{H}_b  \rho \rangle \approx0,
\end{equation}
and similarly for the gauge transformation generated by the Hamiltonian. We can seemingly continue to do this indefinitely. We come back to this subtlety for the CQ theory at the end of section \eqref{sec: scalarfield}. For now, we will take the consistency conditions in a weak form asking whether or not there exists a sensible set of observables and states for which they hold. 
%%%

\subsection{Dirac argument for CQ master equations}\label{sec: diracmaster}
The Dirac argument extends naturally to the CQ theory. In particular suppose we are given a CQ master equation of the form
\begin{align}
\frac{\partial \cqstate}{\partial t} = \mathcal{L}_0( \cqstate) + \lambda^a(t) \mathcal{L}_a( \cqstate),
\end{align}
where the $\lambda^a(t)$ are undetermined functions of time. We can run the Dirac argument on $\cqstate$. Without loss of generality, consider having an initial CQ state $\cqstate(z,0)$ at $t=0$. The master equation depends on some arbitrary functions of time $\lambda^a(t)$. Consider picking two different functions $\lambda^a, \lambda^{a \prime} $. Then at time $t= \epsilon$ the solutions are
\begin{align}
& \cqstate(\epsilon) = \cqstate(0) + \epsilon \mathcal{L}_0 (\cqstate(0)) + \epsilon \lambda^a \mathcal{L}_a ( \cqstate) \\
& \cqstate'(\epsilon) = \cqstate(0)+ \epsilon \mathcal{L}_0 (\cqstate(0)) + \epsilon \lambda^{a \prime} \mathcal{L}_a ( \cqstate(0)).
\end{align}
The difference between the two solutions is 
\begin{align}
\delta \cqstate(\epsilon) = \epsilon (\lambda^{a \prime } - \lambda^a) \mathcal{L}_a( \cqstate(0)),
\end{align}
from which we conclude that $\mathcal{L}_a ( \cqstate)$ is generating equal time gauge transformations. We should then ask that the generators close as an algebra.
\begin{align}
\mathcal{L}_a ( \mathcal{L}_b(\cqstate))-\mathcal{L}_b ( \mathcal{L}_a(\cqstate)) \approx C^{c}_{ab} \mathcal{L}_c(\cqstate).
\label{eq : alg}
\end{align}
Since the classical part of the CQ state may only be defined in a distributional sense, we can re-state \eqref{eq : alg} as asking that
\begin{equation}
   \langle A, \mathcal{L}_a ( \mathcal{L}_b(\cqstate))-\mathcal{L}_b ( \mathcal{L}_a(\cqstate)) \rangle \approx 0
\end{equation}
whenever $A$ is an observable. Similar to the case of classical GR in the previous subsection, we will see that in CQ theories of gravity the generators only close as an algebra if we are on a constraint surface. We will ask that the constraints are satisfied in the weakest sense. That is, we will look for a non-trivial set of states and observables for which the consistency condition is satisfied and preserved in time. Using this method we are able to derive a generalized momentum constraint for the CQ theory; preservation of this constraint will lead to a generalized Hamiltonian constraint.

\section{Deriving constraints in post-quantum theories of Gravity}\label{sec: constraintfull}
In this section we will use the Dirac argument to derive consistency conditions for CQ theories of gravity. We expect the methods used here to extend to all CQ theories of the form introduced in section \ref{sec: CQeinstein}, namely those which are linear in the lapse and shift. We briefly outline a general procedure for generating consistency conditions and constraints in such theories. We spend the remainder of the paper performing the explicit calculations for a specific class of CQ theories of a quantum scalar field coupled to gravitational degrees of freedom.

\subsection{A general method of arriving at constraints}\label{sec: genmethod}
We consider CQ dynamics linear in the lapse and shift which reproduce Hamiltonian evolution in the classical limit. In particular we will consider we consider the simplified theory of equation \eqref{eq:dynamicalreal}
\begin{align}
\frac{\partial\cqstate(g,\pi)}{\partial t}=
\int d^3x {N \mathcal{L}(\cqstate)} + { N^a \mathcal{L}_a (\cqstate) } = \mathcal{L}[N] + \mathcal{L}[\vec{N}],
\label{eq:dynamicalPQG}
\end{align}
where
\begin{align}
&\begin{split}
\mathcal{L}(x)(\cqstate)  &= \{ \ch (x) , \cqstate \} -i[ \qh(x), \cqstate] \\
& + W^{ \ag \bg}(x)\lin_\ag(x)\cqstate\lin_\bg^\dagger(x) - \frac{1}{2 } W_0^{\ag \bg}(x)\{\lin_\bg^\dagger(x)\lin_\ag(x),\cqstate\}
\end{split}\\
&\begin{split}
\mathcal{L}_a(x)(\cqstate) &= \{ \cm_a(x), \cqstate \} -i[ \qm_{m,a}(x), \cqstate] \\
& + W_a^{\ag \bg}(x)\lin_{\ag}(x) \cqstate \lin_{\bg}^{\dag}(x) - \frac{1}{2 } W_{0 a}^{\ag \bg}(x)\{\lin_\bg^\dagger(x)\lin_\ag(x),\cqstate\}.
\end{split}
\label{eq:gener}
\end{align}
The Dirac argument tells us that solutions related by different choices of $N,N^a$ must be gauge equivalent, which tells us that $\mathcal{L}(x), \mathcal{L}_a(x)$ generate equal time gauge transformations. 

As a consequence, denoting the dual operators to $\mathcal{L}$ and $\mathcal{L}_a$ as $\mathcal{L}^*, \mathcal{L}^*_a$ respectively, this implies that 
\begin{equation}
\int Dg D\pi \mathcal{L}^*( A) \cqstate \approx 0, \s \int Dg D\pi \mathcal{L}^*_a( A) \cqstate \approx 0
\end{equation} for all observables $A$.\footnote{Here, the dual operators are defined via integration by parts $\int D g D \pi A \mathcal{L}(\cqstate) = \int Dg D \pi \mathcal{L}^*(A) \cqstate$.} In section \ref{sec: scalarfield} we will only consider realizations which diffuse in the conjugate momenta $\pi^{ab}$. Hence, we expect that any constraints will be solved by functions of the form $\cqstate(g, \pi) = \delta(g,\bar{g}) \cqstate_c(g, \pi)$, since the commutation relations will never lead to $g_{ab}$ diffusive terms. As a consequence, we expect that 
\begin{equation}
\int D\pi \mathcal{L}^*( A)( \bar{g}, \pi) \cqstate_c(\bar{g}, \pi) \approx 0, \s \int D\pi \mathcal{L}_a^*( A)( \bar{g}, \pi) \cqstate_c(\bar{g}, \pi) \approx 0.
\label{eq: gauge2}
\end{equation}
Multiplying equation \eqref{eq: gauge2} by an arbitrary function of the metric $N(\bar{g})$, putting back the delta $\delta(g,\bar{g})$, and integrating over $g$, we expect more generally that 
\begin{equation}
    \int Dg D\pi \mathcal{L}^*( A) N(g)  \cqstate \approx 0, \s \int Dg D\pi \mathcal{L}^*_a( A) N(g)\cqstate \approx 0,
\end{equation}
so that the gauge transformation can have a metric dependence on the lapse and shift.
The requirement that the gauge transformations are transitive leads to consideration of the algebra of generators\footnote{For the theory considered in section \ref{sec: scalarfield} one can verify that the consideration of the algebra in \eqref{alg} is equivalent to the algebra generated by $ \mathcal{L}[N], \mathcal{L}[\vec{N}]$, where $N= N(t,x,g)$ is a functional of the metric; the difference is a metric dependent gauge transformation. However, more generally one would have to consider the algebra which included the metric dependent generators.}
\begin{align}
[ \mathcal{L}_a(x), \mathcal{L}_b(y)], \s [ \mathcal{L}_a(x), \mathcal{L}(y)], \s [ \mathcal{L}(x), \mathcal{L}(y)],
\label{alg}
\end{align}
where once again we remind the reader that $\mathcal{L}, \mathcal{L}_a$ are classical-quantum operators acting on $\cqstate$, in the sense of equation \eqref{eq:dynamicalPQG}, where the couplings $W^{\alpha \beta}(x)$ are to be interpreted as kernels $ W^{\ag \bg}(x)(\cqstate) = \int d z' W^{\ag \bg} ( z | z',x) \cqstate(z') $, and similarly for $W_a^{\alpha \beta}(x)$.  
For example, $[ \mathcal{L}_a(x), \mathcal{L}_b(y)](\cqstate)=\mathcal{L}_a(x)(\mathcal{L}_b(y)(\cqstate))-\mathcal{L}_b(x)(\mathcal{L}_a(y)(\cqstate)).$

We require that the algebra closes; that is the commutator of two gauge transformations is weakly another gauge transformation, which vanishes when smeared over an observable $A \in O_{obs}$. In a similar fashion to GR we will see that demanding this will lead to a notion of a constraint surface. We will often deal with the spatially smeared versions of these transformations and denote $\mathcal{L}[N] = \int d^3 x N(x) \mathcal{L}(x) $ and $\mathcal{L}[\vec{N}] = \int d^3 x N^a(x) \mathcal{L}_a(x) $.

\subsection{A CQ theory of gravity coupled to a scalar field}\label{sec: scalarfield}
Now that we have a general method of deriving constraints in CQ theories, we shall spend the remainder of the paper exploring the consistency conditions for a quantum scalar field interacting with a classical gravitational field. 
Many of the considerations, however, are more general, although we make a number of simplifying 
assumptions so that the theory considered here is a special case of the one derived in \cite{oppenheim_post-quantum_2018}.
\begin{assumption}\label{as:linear}
    We take the evolution to be linear in the lapse and shift, so that $\mathcal{L}[N] = \int d^3 x N(x) \mathcal{L}(x) $ 
    and consider the choice of $N$ and $\vec{N}$ to be pure gauge.
    \label{as:gauge}
\end{assumption}
\begin{assumption}\label{as:spatialDiff}
    We will take the $W^{ \ag \bg}_a(x)=0$, so that $\mathcal{L}_a( \cqstate) = \{ \cm_a(x), \cqstate \} -i[ \qm_{m,a}(x), \cqstate] $ generates spatial diffeomorphisms and the theory will be spatially diffeomorphism invariant.
\end{assumption}

\begin{assumption} \label{as:1}
  We take the CQ couplings $W^{\ag \bg}(z|z') $ to have the same Lindblad structure as the Hamiltonian so that the interaction terms can be written 
\begin{align}\nonumber
 & W^{ \ag \bg}(x)\lin_\ag(x)\cqstate\lin_\bg^\dagger(x) = W^{\phi \phi}(x) \phi(x) \cqstate \phi(x) +  W^{\pi \pi}(x) \pi_\phi(x)  \cqstate \pi_\phi(x) \\ 
 & + W^{ab}(x)\partial_a \phi(x) \cqstate \partial_b \phi(x) .
 \end{align}
In particular,  $W^{ \ag \bg}(x)$ is taken to be local in $x$, while the more general case could be non-local and include a regulator $W^{ \ag \bg}(x-y)\lin_\ag(x)\cqstate\lin_\bg^\dagger(y)$.
\end{assumption}

\begin{assumption}\label{as:2}
  We take the CQ coupling with the scalar field, in analogy with classical gravity, to have a functional dependence on the spatial metric only; $W^{\ag \bg} [g](x)$. We call such a coupling  \textit{minimal coupling}.
\end{assumption}
\begin{assumption}
    We take the first moment of $W^{ \ag \bg}(x)$ to reduce to General Relativity in the classical limit, so that $\Tr{W^{ \ag \bg}(x)\lin_\ag(x)\cqstate\lin_\bg^\dagger(x)}=\Tr{\{H_m,\cqstate\}}$  as discussed in Section \ref{sec: CQham}. Since one generally considers matter Hamiltonians which only depend on $g_{ij}$ and not $\pi^{ij}$, this motivates our previous assumption of minimal coupling.
\end{assumption}
\begin{assumption}\label{as:3}
   The CQ term couples states with different momenta $\pi^{ab}$ only; we only jump in momentum. We call such theories $\pi-dispersive$, while in the more general case one can have both dispersion in the $\pi^{ij}$ and dispersion in $g_{ij}$. If both terms are present, then the relationship between $\pi^{ij}$ and $\dot{g}_{ij}$ exists only on the level of expectation values.
\end{assumption}
\begin{assumption}\label{as:pure}
    We take the pure gravity part of the master equation to be deterministic, and given by general relativity $\{H_{ADN},\cqstate\}$. In the more general case the pure gravity evolution can be stochastic.
\end{assumption}

Realisations which satisfy these assumptions are given in appendix \ref{sec: realisations}. 
Assumption \ref{as:gauge} is respected in Einstein's theory of General Relativity, but Einstein also considered
what is now known as the unimodular theory of gravity \cite{einstein1952gravitational,van1982exchange,weinberg1989cosmological,unruh1989unimodular,alvarez2005can,smolin2009quantization, shaposhnikov2009scale} where
$N$ is chosen so that the cosmological constant becomes an integration constant. Assumption \ref{as:gauge} also
doesn't hold in
 Horava gravity \cite{horava, horavaphase,horavaphase2} and shape dynamics \cite{mercati2017shape, Anderson_2005, Gomes_2012}, but here we explore the consequences of taking the full gauge symmetry.
Assumption \ref{as:1} seems reasonable to ask: essentially, when computing the commutators of the gauge transformations it must be the case that $W^{\ag \bg}$ term must transform like the Hamiltonian, or else the pure classical part and pure quantum part of the evolution will transform differently to the CQ coupling. One could imagine a theory with alternative, higher order, Lindblad operators which have the correct transformation properties -- such as $ W^{\phi \phi \phi \phi}(x)\phi \phi \cqstate \phi \phi$ -- but we do not discuss such theories in this paper and leave this type of realization as a possible alternative. 

We can summarize assumptions \ref{as:2} and \ref{as:3} as considering CQ couplings which take the form $W^{\ag \bg}(z|z',x) = W^{\ag \bg}(g,\pi|g', \pi',x)$, where the moments of $ W^{\ag \bg}(g,\pi|g, \pi',x)$ depend only on $g_{ab}$ and not $\pi^{ab}$. The assumptions we make here are motivated two fold. Firstly, by analogy to pure classical gravity. There, when one considers minimally coupled scalar fields the interaction term is of the form 
\begin{equation}
W^{\alpha \beta}_{classical}(x)( \rho) = \int d^3 y \frac{\delta \mathcal{H}_m(x)}{\delta g^{ab}(y)}  \frac{\delta \rho}{\delta \pi^{ab}(y)} = \int d^3 y \frac{\delta h^{\alpha \beta}(x)}{\delta g^{ab}(y)} L_{\beta}^{\dag} L_{\alpha} \frac{\delta \rho}{\delta \pi^{ab}(y)} ,
\end{equation}
which only couples states with different momenta, and the coupling only has a functional dependence on the spatial metric through $\frac{\delta h^{\alpha \beta}}{\delta g^{ab}}$. Secondly, since we end up calculating the commutation relations in \eqref{alg}, which includes Poisson brackets with the pure classical Hamiltonian and momentum, assumptions \ref{as:1}, \ref{as:2}, \ref{as:3} seem natural. If, for example, one has diffusion in the spatial metric then one quickly runs into very messy calculations. However, we emphasise that we do not have any evidence such theories could not be consistent and leave this as a possible direction for future research.

Having made these simplifying assumptions, we are now ready to study their constraints. We have not specified the CQ coupling explicitly, except for some Lindbladian structure and functional dependence that we would like it to have. As we shall see in the next section, there are various transformation properties that any realization must satisfy. We then study the constraints in such realizations, which arises from the study of the algebra of equal time gauge generators in \eqref{alg}. In particular, we find that the $ [ \mathcal{L}_a(x), \mathcal{L}_b(y)]$ commutator closes, which is an expected artifact of the fact we are assuming (assumption \ref{as:spatialDiff}) $\mathcal{L}_a$ generates spatial diffeomorphisms. We find the $[ \mathcal{L}_a(x), \mathcal{L}(y)]$ generator closes so long as the couplings $W^{\ag \bg}(z|z',x)$ satisfy certain transformation rules; essentially telling us that $W^{\ag \bg}(z|z',x)$ must transform correctly under spatial diffeomorphisms . Finally, we study the $[ \mathcal{L}(x), \mathcal{L}(y)]$ commutator. This will lead to a constraint which we interpret as a generalization of the momentum constraint to CQ dynamics. We find preservation of this constraint gives rise to a CQ analog of the Hamiltonian constraint. 
\subsubsection{\texorpdfstring{$[ \mathcal{L}_a(x), \mathcal{L}_b(y)]$ commutator for the scalar field}{what}}\label{sec: momcom}
Due to assumption \ref{as:spatialDiff}, we are considering the case where $W^{\alpha \beta}_a =0$, so that the total CQ momentum generator $\mathcal{L}_a$  reads
\begin{equation}
\mathcal{L}_a(x)(\cqstate) = \{ \cm_a(x), \cqstate \} -i[ \qm_{m,a}(x), \cqstate] .
\label{momf}
\end{equation}
One can then easily verify that 
\begin{equation}
[\mathcal{L}[\vec{N}], \mathcal{L}[ \vec{M}]]  = \mathcal{L}[ [\vec{N}, \vec{M}]],
\label{momalg}
\end{equation}
which vanishes when smeared over an observable since $\mathcal{L}_a$ is treated as gauge. This verifies that \eqref{momf} is the generator of spatial diffeomorphisms.

\subsubsection{ \texorpdfstring{$[ \mathcal{L}_a(x), \mathcal{L}(y)]$}{what} commutator for the scalar field }\label{sec: momham}
We now compute the $[ \mathcal{L}_a(x), \mathcal{L}(y)]$ commutator. With the interpretation that $\mathcal{L}_a$ generates spatial diffeomorphisms we find that the algebra closes so long as we pick the realizations to have the correct transformation  properties under spatial diffeomorphisms. In total we find the (smeared) commutation relation
\begin{equation}\label{momham}
\begin{split}
& [ \mathcal{L}[\vec{N}], \mathcal{L}[N] ](\rho) = \int d^3 x \left[N^a D_a N \mathcal{L}(\rho)\right] \\
& + \int d^3 x NN^a\left[ L_{\ag} D_a(W^{\ag \bg}\cqstate) L_{\bg}^{\dag} - \frac{1}{2} \{  L_{\bg}^{\dag} L_{\ag}, D_a(W_0^{\ag \bg}\cqstate) \} \}\right] \\ 
& +\int d^3 x  D_a N^a \left[  W^{\phi \phi} \phi \cqstate \phi - \frac{1}{2}\{ W_0^{\phi \phi} \phi^2, \cqstate \} \right] 
-  D_a N^a \left[  W^{\pi \pi} \pi \cqstate \pi - \frac{1}{2}\{ W_0^{\pi \pi} \pi^2, \cqstate \} \right]\\
& +  \int d^3 x \left[(D_c N^c W^{ab} - D_c N^b W^{ca}  - D_c N^a W^{cb}  ) (D_a \phi \cqstate D_b \phi - \frac{1}{2} \{ D_b \phi D_a \phi, \cqstate \} ) \right] \\ 
& +  \int d^3 x N(x) \left[ \{ H[ \vec{N}], W^{\ag \bg} L_{\ag} \cqstate L_{\bg}^{\dag} - \frac{1}{2} W_0^{\ag \bg} L_{\bg}^{\dag} L_{\ag}, \cqstate \} \}\right. \\
&  - \left.  W^{\ag \bg} L_{\ag} \{ H[\vec{N}], \cqstate \}  L_{\bg}^{\dag} - \frac{1}{2} W_0^{\ag \bg} L_{\bg}^{\dag} L_{\ag}, \{ P[\vec{N}], \cqstate \} \} \right].
\end{split}
\end{equation}
Although somewhat daunting, we see that so long as $W^{ \ag \bg}$ satisfies certain transformation properties, specifically if  \begin{equation}\label{eq: Hojm}
\begin{split}
&\{ \cm[ \vec{N}] , W^{\pi \pi} (\cqstate) \} - W^{\pi \pi} ( \{ \cm[ \vec{N}], \cqstate \} ) =  \int d^3 x D_a N^a  W^{ \pi \pi}( \cqstate)  - N^a D_a W^{\pi \pi}(\cqstate), \\ 
&\{ \cm[ \vec{N}] , W^{\phi \phi} (\cqstate) \} - W^{\phi \phi} ( \{ \cm[ \vec{N}], \cqstate \} ) =  \int d^3 x -D_a N^a W^{ \phi \phi}( \cqstate)  -N^a D_a W^{\phi \phi}(\cqstate),\\ 
& \{ \cm[ \vec{N}] , W^{a b} (\cqstate) \} - W^{a b } ( \{ \cm[ \vec{N}], \cqstate \} ) \\
& =  \int d^3 x [D_c N^b W^{ca} (\cqstate)  + D_c N^a W^{cb} (\cqstate) - D_c N^c W^{ab} (\cqstate) ] - N^a D_a W^{a b }(\cqstate) ,
\end{split}
\end{equation}
then the algebra will close
\begin{align}
[ \mathcal{L}[\vec{N}], \mathcal{L}[N] ](\rho) = \int d^3 x N^a D_a N \mathcal{L}(\rho)  = \mathcal{L}[ L^{\text{Lie}}_{\vec{N}} M] (\rho),
\end{align}
and the theory will be spatially diffeomorphism invariant. Here $ L_{\vec{N}} N $ is the Lie derivative of $N$ along $N^a$. 
The conditions in equation \eqref{eq: Hojm} are demystified somewhat when one realises they are the analogous to terms arising in classical gravity. It is a general property of minimally coupled field theories that \cite{hojman1976geometrodynamics}
\begin{align}\label{eq: hojy}
& \{ \mathcal{H}_{ma} (x) , \mathcal{H}_m(y) \} = 2 D_b\left(\frac{\delta \mathcal{H}_m(y)}{\delta g_{ab}(x)}\right) + \mathcal{H}_m(x) \partial_a \delta(x,y)\\
& \{ \mathcal{H}_a(x), \mathcal{H}_m(y) \} = -2 D_b\left(\frac{\delta \mathcal{H}_m(y)}{\delta g_{ab}(x)}\right),
\label{eq: hojm}
\end{align}
and these combine so that the matter Hamiltonian transforms as a scalar under spatial diffeomorphisms 
\begin{equation}
\{\mathcal{H}_a + \mathcal{H}_{ma}(x), \mathcal{H}_m(y) \} = \mathcal{H}_m(x) \partial_a \delta(x,y).
\end{equation} Since we are assuming that the CQ Lindbladian has the same structure as the matter Hamiltonian, we often find terms which look similar to \eqref{eq: hojy} and \eqref{eq: hojm} -- essentially the anomalous terms in \eqref{momham}. We then expect these to cancel with terms arising from the Poisson bracket -- the terms in the final line of \eqref{momham} -- enforcing conditions on the allowed realizations. We interpret this as telling us that $W^{\ag \bg}(z,z')L_{\bg}^\dagger L_\ag$ must transform like the Hamiltonian under the action of the Poisson bracket. 
From now on we will assume that these can be satisfied, without reference to an explicit realization and we will derive the constraints which arise from the final component of the algebra -- the $[ \mathcal{L}(x), \mathcal{L}(y)]$  commutator. We take up the construction of a realisation in appendix \ref{sec: realisations}.

\subsubsection{\texorpdfstring{$[ \mathcal{L}(x), \mathcal{L}(y)]$ commutator for the scalar field}{hm}}\label{sec: hamham}
We now move on to study the final commutator in the algebra. So far we have found restrictions on the realizations of the CQ theory. In this section, we will find that we will need to impose constraints in order for the theory to be consistent.  A long calculation yields
\begin{align}
[\mathcal{L}[N], \mathcal{L}[M] ](\cqstate)=  \int d^3 x (N \partial_a M - M \partial_a N) \left[ \{ g^{ab} \cm_b, \cqstate \} - i[ g^{ab} \qm_{m,b}, \cqstate] + \bar{\mathcal{C}}^a(\cqstate) \right],
\label{eq: hamham}
\end{align}
where 
\begin{align}
\bar{\mathcal{C}}^a (\cqstate) & =  \con_J^{ab} ( D_b \bphi(x) \cqstate \bpi(x) + \bpi(x) \cqstate D_b \bphi(x) ) - \frac{1}{2} \con^{ab}_N \{ D_b \phi(x) \pi(x) + \pi(x) D_b \phi(x) , \cqstate \}_+ \nonumber \\
& -i\con_H^{ab} [ \qm_a, \cqstate]  + i\con_{JN}^{ab} ( D_a \bphi(x) \cqstate \bpi(x) - \bpi(x) \cqstate D_a \bphi(x)).
\label{eq: constraint}
\end{align}
In \eqref{eq: constraint} we have defined the couplings\footnote{Here the labels $J,N,H,JN $ stand for ``Jump", ``no-event", ``Hamiltonian", ``jump-no-event", which follows from the convention adopted just after equation \eqref{eq: cqdyngen}. Specifically, the $C_J^{ab}$ term takes the form of a Jump term in a Lindblad equation, whilst $C_N^{ab}$ takes the form of a no-event term in the Lindblad equation. The $C_H^{ab}$ term takes the form of a Hamiltonian term. The final term $C_{JN}^{ab}$ has no analogue with the Lindblad equation, but arises due to the commutation of the jump and no-event term in $\mathcal{L}$.}
\begin{align}
& \con_J^{ab} =2(h^{ab} W^{\pi \pi} +  h^{\pi \pi} W^{ab}), \s \con^{ab}_N= 2(W^{ab}_0h^{\pi \pi} + W^{\pi \pi}_0 h^{ab}) \nonumber \\ & \con^{ab}_H = \frac{1}{4}(W^{\pi \pi} W^{ ab} - W^{ab}_0 W^{\pi \pi}_0) , \s \con^{ab}_{JN} = ( W^{ab} W_0^{\pi \pi} - W^{ab}_0 W^{\pi \pi}) .
\label{eq: definitions}
\end{align}
The first two terms of equation \eqref{eq: hamham} give rise to a component which is a gauge transformation. To see this we smear over a phase space test function $A$ to find
\begin{equation}
\begin{split}
\langle A, \{ g^{ab} \mathcal{H}_b, \cqstate \} -  i[ g^{ab} \qm_{m,b}, \cqstate] \rangle  &=  \langle \{ A, g^{ab} \}, \mathcal{H}_b  \cqstate \rangle +  \langle \{ A, \mathcal{H}_b \}g^{ab},  - i[ g^{ab} \qm_{m,b}, \cqstate] \rangle \\
& = \langle \{ A, g^{ab} \}, \mathcal{H}_b  \cqstate \rangle + \langle \mathcal{L}_b^*(A) g^{ab}, \cqstate \rangle  ,
\label{eq: calc}
\end{split}
\end{equation}
which, since the final term in \eqref{eq: calc} is a gauge transformation, is weakly equal to
\begin{equation}
 \langle \{ A, g^{ab} \}, \mathcal{H}_b  \cqstate \rangle.
\end{equation}
As a consequence, we find 
\begin{align}
[\mathcal{L}[N], \mathcal{L}[M] ](\cqstate) \approx  \int d^3 x (N \partial_a M - M \partial_a N) \left[ \{ g^{ab}, \mathcal{H}_b \cqstate \} + \bar{\mathcal{C}}^a( \cqstate)\right],
\label{eq: hamham2}
\end{align}
and so in order for the algebra of generators to close we need to impose the CQ momentum constraint
\begin{align}
\mathcal{C}^a = \{ g^{ab}, \mathcal{H}_b \cqstate \} + \bar{\mathcal{C}}^a( \cqstate) \approx 0,
\end{align}
which should hold when smeared over observables $A$. Using the definitions in \eqref{eq: definitions}, we can write this out in full as
\begin{align}\label{eq: momentumConstraintFull}
\mathcal{C}^a (\cqstate) & = \{ g^{ab}, \cm_b \cqstate \}-iC_H^{ab}[ \qm_{m,a}, \cqstate] + C_J^{ab} ( D_b \bphi(x) \cqstate \bpi(x) + \bpi(x) \cqstate D_b \bphi(x) )   \nonumber \\
& -\frac{1}{2} C_N^{ab} \{ \qm_{m,a}, \cqstate \}_++ C_{JN}^{ab}( D_a \bphi(x) \cqstate \bpi(x) - \bpi(x) \cqstate D_a \bphi(x)).
\end{align}

It is useful to perform a quick sanity check on equation \eqref{eq: momentumConstraintFull} to see if it gives the correct momentum constraint in the classical limit. First, taking the trace over the quantum system and defining $\Tr( \cqstate(z)) = \rho(z)$ one sees
\begin{equation}
\Tr( C^a(\cqstate) ) = \{ g^{ab}, \cm_b \rho \} + \Tr{}{\left[(2C_J^{ab}- C_N^{ab})  D_b \phi \pi \cqstate\right]}.
\label{eq: trace}
\end{equation}
Then, performing a Kramers-Moyal expansion of the CQ couplings $W^{\ag \bg}(z|z',x)$ to first order, one finds the zeroth order term cancels in equation \eqref{eq: trace} and, remembering that the first moment is such that Einstein's equations hold, we are left with
\begin{equation}
\Tr( C^a(\cqstate) ) = \{ g^{ab}, \cm_b \rho \} + \Tr(D_b \phi \pi \{ g^{ab}, \cqstate\} ) + \dots \ .
\label{eq: tracemomkramers}
\end{equation}
Smearing over an observable $A$, equation \eqref{eq: tracemomkramers} becomes 
\begin{equation}
\int Dg D\pi A \Tr( C^a(\cqstate) ) =  \int Dg D\pi \{A, g^{ab}\}( \cm_b \rho +  \Tr(D_b \phi \pi \cqstate))+ \dots \ .
\label{eq: classham}
\end{equation}
Comparing this CQ constraint to the standard momentum constraint of GR, we see that it gives a sensible constraint in the classical limit; namely one which is satisfied by classical gravity. We also see that in the limit where the matter remains quantum, we get a sensible constraint, in the sense that it appears satisfiable even though it contains both functionals of the classical degrees of freedom, and quantum operators. In particular, one could have been concerned that we would find that we had to satisfy the naive CQ constraint $\mathcal{H}_a(x)+\qm_{m,a}(x)\approx 0$ when restricted to the constraint surface. This would have required setting a c-number equal to an operator equation. Instead, we get a CQ-equation that appears to be equivalent to finding mixed fixed points of some dynamics, which is at least possible to hold in principle.

\subsubsection{Conservation of the momentum constraint \texorpdfstring{$C^a$}{C} for the scalar field }\label{sec: momcon}
Now that we have a momentum constraint, we must check to see if it preserved in time. In the classical case, preservation of the momentum constraint gives rise to the Hamiltonian constraint and we expect something analogous for the CQ theory. Indeed, we shall find we get a constraint which is the standard Hamiltonian constraint in the classical limit -- although it appears to require additional constraints or a restriction on the lapse and shift if one is to hope that it will be preserved in time. We discuss the possible implications for this when we conclude in section \ref{sec: discussion}.

Conservation of the momentum constraint requires calculating the quantity $C^a(\frac{\partial \cqstate }{\partial t})= C^{a}( \mathcal{L}[N,\vec{N}](\cqstate)$. For calculation purposes, it is slightly simpler to consider the commutator $[C^a,\mathcal{L}[N,\vec{N}]](\cqstate)$, noting that difference between the commutator and the evolution of the constraint is given by the term $\int Dg D\pi C^{a*}( \mathcal{L}^*(A))\cqstate$; that is the momentum constraint but smeared over the phase space operator $\mathcal{L}^*(A)$ instead of $A$. When performing the calculation, we will smear the momentum constraint  $C^a$ with a lower index spatial smearing function $M_a(x)$ and write $C^a$  $\mathcal{C}[ \underbar{M}] = \int d^3 x M_a \mathcal{C}^a $.

We first calculate the commutator with the spatial part of the evolution equation $[C[\underbar{M}], \mathcal{L}[\vec{N}]](\cqstate)$; in other words to check whether or not the momentum constraint transforms correctly under spatial diffeomorphisms. Assuming that the realization has the transformation properties defined in  \eqref{eq: Hojm}, one finds\footnote{This can be made to look similar to the $\mathcal{L}_a, \mathcal{L}_b $ commutator by using integration by parts on the second term of \eqref{eq: momcons} where it differs slightly because $D_b C^a(\cqstate)$ is not vanishing.}
\begin{align}
 [\mathcal{C}[\underbar{M}], [\mathcal{L}[ \vec{N}]](\cqstate) = -\int d^3 x N^c D_c M_a \mathcal{C}^b( \cqstate) + M_a D_c N^a \mathcal{C}^c (\cqstate) ,
\label{eq: momcons}
\end{align}
which vanishes on the constraint surface. We are then left to calculate the commutation with $\mathcal{L}[ N]$, which, in analogy to the classical case, we expect to give a Hamiltonian constraint.

It shall be useful to split up the generators into the purely classical part involving the Poisson bracket and everything else. To that end, we write $\mathcal{L}(\cqstate) = \{ \ch, \cqstate \} + \bar{\mathcal{L}}(\cqstate)$ and $C^a(\cqstate) = \{ g^{ab}, C^b \cqstate \} + \bar{C}^a(\cqstate)$. The evolution of the smeared constraint reads
\begin{equation}
\begin{split}
[C[\underbar{M}], \mathcal{L}[N]](\cqstate)=&  \int d^3 x d^3 y M_a(y) N(x) \big [ \{ g^{ab}(y), \{ \mathcal{H}_b(y),\ch(x) \} \cqstate \}  \\
& +  \{ g^{ab}, \cm_b \bar{\mathcal{L}}(\cqstate) \}-\bar{\mathcal{L}}( \{ g^{ab}, \cm_b \cqstate \}) + [ \bar{\mathcal{C}}^a, \bar{\mathcal{L}}](\cqstate)  ) \\
&+ \{\{ g^{ab}(y), \mathcal{H}(x)\}, \cm_b(y) \cqstate \} +  \bar{\mathcal{C}}^a ( \{ \ch, \cqstate \} -\{ \ch, \bar{\mathcal{C}}^a(\cqstate) \} \big].
\label{eq: hamcon}
\end{split}
\end{equation}
Now there is a lot going on here, so we will break \eqref{eq: hamcon} into pieces and discuss what we expect to get back from each term, before presenting our findings. We first comment on the third line of equation \eqref{eq: hamcon}, which consists of the term
\begin{equation}
 \{\{ g^{ab}(y), \mathcal{H}(x)\}, \cm_b(y) \cqstate \} +  \bar{\mathcal{C}}^a ( \{ \ch, \cqstate \} -\{ \ch, \bar{\mathcal{C}}^a(\cqstate) \}.
 \label{eq: sbm}
\end{equation}
Firstly, we note that \eqref{eq: sbm} has the Lindblad structure of the momentum constraint.\footnote{By this we mean that the expression in \eqref{eq: sbm} contains quantum operators with the same structure as the momentum constraint \eqref{eq: momentumConstraintFull}, i.e those which come in the format $\sim \partial_a \phi \cqstate \pi$.} To gain some intuition it is useful to take the trace of \eqref{eq: sbm} and look at first order in the Kramers-Moyal expansion. Explicitly, equation \eqref{eq: sbm} becomes
\begin{equation}
 \{\{ g^{ab}(y), \mathcal{H}(x)\}, \cm_b(y) \rho \} + \Tr( D_b \phi \pi \{ \{ g^{ab}(y), \ch(x) \}, \cqstate \})  + \dots
\end{equation}
and smearing this over an observable gives
\begin{equation}
 \int Dg D\pi \{A, \{ g^{ab}(y), \mathcal{H}(x)\}\}( \cm_b \rho +  \Tr(D_b \phi \pi \cqstate))+ \dots \ .
 \label{eq: tracemom}
\end{equation}
We can compare \eqref{eq: tracemom} to the momentum constraint in \eqref{eq: classham} and we note that although almost identical they are not quite the same. To be precise, equation \eqref{eq: tracemom} is smeared over $\{ A, \{ g^{ab}, \mathcal{H} \}\}$ whilst equation \eqref{eq: classham} is instead smeared over $\{ A, g^{ab} \} $. Of course, both the constraints are not independent; they both vanish if we satisfy the effective classical constraint $\cm_b \rho +  \Tr(D_b \phi \pi \cqstate) =0$. We therefore posit that in a sensible realization equation \eqref{eq: sbm} should be weakly zero whenever the momentum constraint is satisfied. Otherwise, one is forced to view equation \eqref{eq: sbm} as a separate constraint to the momentum constraint, which must be preserved in time by itself -- one then faces a similar issue looking at the time evolution of equation \eqref{eq: sbm} and must impose more constraints in a series which seems unlikely to terminate. We view the condition that equation \eqref{eq: sbm} should be weakly zero whenever the momentum constraint is satisfied as a transformation rule which any realization $W^{\ag \bg}(z|z')$ must obey, telling us how they CQ couplings must transform under the action of the pure gravity Hamiltonian $\{ \ch, \}$ -- just as the transformation rules defined in equation \eqref{eq: Hojm} place conditions on the realizations in order for the theory to be spatially diffeomorphism invariant. 

We now study the remaining terms in equation \eqref{eq: hamcon}, given by the first two lines,  which we expect to give rise to a Hamiltonian constraint. Before presenting the result, it is perhaps useful to get some intuition for each of the terms which appear. Using the Dirac algebra, defined in \eqref{eq: defalg}, the purely classical term, $\{ g^{ab}(y), \{ \mathcal{H}_b(y),\ch(x) \} \cqstate \}$, can be written as $\partial_b^y \delta(y,x)\{ g^{ab}(y), \ch(y)\cqstate \} $, giving rise to the classical Hamiltonian part of the constraint. 
It is less obvious what we expect $[ \bar{\mathcal{C}}^a, \bar{\mathcal{L}}](\cqstate)$ to give back. Taking a step back and looking at the analogous term in classical gravity, one has instead of $[ \bar{\mathcal{C}}^a, \bar{\mathcal{L}}](\cqstate)$ the term
\begin{equation}
\{ \mathcal{H}_{ma} (x) , \mathcal{H}_m(y) \} = 2 D_b\left(\frac{\delta \mathcal{H}_m(y)}{\delta g_{ab}(x)}\right) + \mathcal{H}_m(x) \partial_a \delta(x,y)
\end{equation}
which gives rise to the Hamiltonian constraint and a term $2 D_b\left(\frac{\delta \mathcal{H}_m(y)}{\delta g_{ab}(x)}\right)$. In calculating the full Poisson bracket between the momentum constraint and the Hamiltonian constraint this anomalous term cancels with a term arising from the Poisson bracket between the pure gravity momentum and the matter Hamiltonian $\{ \mathcal{H}_a, \mathcal{H}_m \} $ so that in combination $\{\mathcal{H}_a+ \mathcal{H}_{ma}, \mathcal{H}_m \} \sim \mathcal{H}_m$ gives back the matter part of the Hamiltonian constraint. 

By virtue of the Lindblad structure of the constraints, due to assumption \ref{as:1}, we expect that under commutation $\bar{C}^a$ transforms like a momentum, whilst $\mathcal{L}$ transforms like a Hamiltonian. As a consequence we expect to find (schematically)
\begin{equation}
[ \bar{C}^a, \bar{\mathcal{L}}] \sim \mathcal{R} + \mathcal{L}_{constraint}(x) \partial_a \delta(x,y),
\label{eq: Cqhoj}
\end{equation}
 where $\mathcal{R}$ is the CQ version of $ 2 D_b\left(\frac{\delta \mathcal{H}_m(y)}{\delta g_{ab}(x)}\right)$ and $ \mathcal{L}_{constraint}(x)$ is the CQ Hamiltonian constraint. Finally, in analogy with the classical case, we expect the $\mathcal{R}$ term appearing in equation \eqref{eq: Cqhoj} to cancel with the Poisson bracket term arising in the second line of \eqref{eq: hamcon}, namely $ \{ g^{ab}, \cm_b \bar{\mathcal{L}}(\cqstate) \}-\bar{\mathcal{L}}( \{ g^{ab}, \cm_b \cqstate \}) $, so that the first two lines of equation \eqref{eq: hamcon} gives rise to the CQ generalization of the Hamiltonian constraint. 
 
We now present the full Hamiltonian constraint -- the first two lines of equation \eqref{eq: hamcon}. In doing so it is first useful to introduce notation for a certain combination of terms which frequently arises. We define $\parts_{AB} $ via
\begin{align}
& \parts_{AB}^{\ag \bg}L_{\ag} \cqstate L_{\bg}=  \int d^3x N\big[ W_B^{\phi \phi} \phi D_b(M_a\con_A^{ab}\cqstate )\phi - W_B^{\pi \pi} \pi  D_b (M_a\con_A^{ab}\cqstate) \pi \nonumber \\  &+ D_e \phi(D_b(M_a\con_A^{af}\cqstate ) W_B^{eb}  
 + D_b(M_a\con_A^{ae}\cqstate ) W_B^{bf} - D_b(M_a\con_A^{ab}\cqstate) W_B^{ef})  D_f \phi  \nonumber \\
 & + M_a C_A^{ab}  L_{\ag} D_b(W_B^{\ag \bg}\cqstate) L_{\bg} \big]
\label{eq: partsdef}.
\end{align}
Here the $A$ sub-index denote terms coming from the momentum constraint and is associated to $C_J,C_H,C_{JN},C_N$, whilst the $B$ sub-index denotes terms coming from the Hamiltonian constraint and is associated to the couplings $W^{\alpha \beta}_J= W^{\alpha \beta}(z|z'), W^{\alpha \beta}_N= W^{\alpha \beta}_0(z), W^{\alpha \beta}_H= h^{\alpha \beta}$.\footnote{Again, the J, H, JN, N stand for jump, Hamiltonian, jump-no event and no-event and are an attempt to label sensibly the terms arising in the constraint. We remind the reader that the jump and no-event terms have the structure defined just before equation \eqref{eq: definitionMoments}, whilst the jump-no event term appearing in the momentum constraint arises from a commutation of jump and no-event terms appearing in the $[\mathcal{L}(x), \mathcal{L}(y)]$ commutator.  } These terms are the CQ analogy of the $ 2 D_b\left(\frac{\delta \mathcal{H}_m(y)}{\delta g_{ab}(x)}\right)$ terms, which arise from integration by parts due the transformation properties of $\bar{\mathcal{C}}^a, \bar{\mathcal{L}}$. A lengthy but straightforward calculation then gives the total Hamiltonian constraint 
\begin{equation}
\begin{split}
& \mathcal{L}_{constraint} = \int d^3 x M_a D_b N \big[ (-2i\con_H^{ab} h^{\ag \bg} + \frac{i}{2}\con_N^{ab} W_0^{\ag \bg} - \frac{i}{2}\con_J^{ab} W^{\alpha \beta}) [L_{\bg}^{\dag}L_{ \ag}, \cqstate ] \\ & +  \{ g^{ab}(y), \ch(y)\cqstate \} +  2( \con_J^{ab} h^{\ag \bg} + \con_H^{ab} W^{\ag \bg})L_{\ag} \cqstate L_{\bg}^{\dag}  - (  \con_N^{ab} h^{\ag \bg} +  \con_H^{ab} W^{\ag \bg}_0) \{ L_{\bg}^{\dag} L_{\ag}, \cqstate \}_+ \big] \\
  & + \int d^3x NM_a \cm_b (\{ g^{ab}, \cm_b \bar{\mathcal{L}}(\cqstate) \}-\bar{\mathcal{L}}( \{ g^{ab}, \cm_b \cqstate \})\\
&+ ( \frac{i}{2} \parts_{NN}^{\ag \bg} - \frac{i}{2}\parts_{JJ}^{\ag \bg} - 2i \parts_{HH}^{\ag \bg} )[ L_{\bg}^{\dag} L_{\ag}, \cqstate] + (2\parts_{HJ}^{\ag \bg} + 2 \parts_{JH}^{\ag \bg}) L_{\ag} \cqstate L_{\bg}^{\dag} \\ 
& -  ( \parts_{NH}^{\ag \bg} + \parts_{HN}^{\ag \bg}) \{ L_{\bg}^{\dag} L_{\ag} \}_+ \\
&+\int d^3 x NM_a\big[( \con_{JN}^{ab}(W_0^{\phi \phi} - 2i h^{\phi \phi}) -\con_{J}^{ab}W^{\phi \phi}_0 +\con_N^{ab} W^{\phi \phi}) D_b \phi \cqstate \phi + \\
& + ( \con_{JN}^{ab}(W_0^{\phi \phi} + 2ih^{\phi \phi}) + \con_{J}^{ab}W_0^{\phi \phi} -\con_N^{ab} W^{\phi \phi})\phi \cqstate D_b \phi \\
&+ ( C_{JN}^{ab}( - W_0^{\pi \pi} +2ih^{\pi \pi}) -W_0^{\pi \pi} \con^{ab}_J + \con_N^{ab} W^{\pi \pi}) D_b\pi \cqstate \pi + ( C_{JN}^{ab}( - W_0^{\pi \pi} -2ih^{\pi \pi}) \\ 
& +( W_0^{\pi \pi} \con^{ab}_J - \con_N^{ab} W^{\pi \pi}) \pi \cqstate D_b \pi \\
&+ ( \con_{JN}^{ab}( -W_0^{\phi \phi} -2ih^{ \phi \phi}) + \con_N^{ab}W^{ef} - W_0^{ef}  h^{ef} \con_J^{ab})( D_e D_f \phi) \cqstate D_b\phi  \\ 
 & + (\con_{JN}^{ab}( -W_0^{\phi \phi} +2ih^{\phi \phi}) - \con_N^{ab}W^{ef} + W_0^{ef} \con_J^{ab})D_b \phi \cqstate (D_e D_f \phi) \big] \\
& \int d^3x \big[  N W^{\pi \pi} \{ \pi D_b(M_a \pi \con_{JN}^{ab} , \cqstate) \}_+ -NM_b\con_{JN}^{ab}W^{\phi \phi} \{ D_b \phi \phi, \cqstate \}_+ \\
& + M_a \con_{JN}^{ab} \{D_b \phi  D_d(NW^{cd} D_c \phi , \cqstate) \}_+ \big]\\
&  -2\int d^3 x \big[  N W^{\pi \pi}_0 \pi D_b( \con_{JN}^{ab} M_a \cqstate)  \pi + M_b D_dN  W^{cd}_0 \con_{JN}^{ab} D_c \phi \cqstate D_b \phi \big] \approx 0 ,
\label{eq: fullham}
\end{split}
\end{equation}
which needs to be weakly zero in order for the theory to be gauge invariant. There is a lot going on so we summarize here was the constraint is telling us.
\begin{itemize}
\item The first two lines of \eqref{eq: fullham} look like a potentially sensible Hamiltonian constraint. Indeed, if we take their quantum trace we end up with \begin{equation}
 \{ g^{ab}(y), \ch(y)\rho \} + 2( C_J^{ab} h^{\ag \bg} + C^{ab}_H W^{\ag \bg} - C_N^{ab} h^{\ag \bg} - C_H^{ab} W_0^{\ag \bg}) \Tr( \lin_{\bg}^{\dag} \lin_{\ag} \cqstate).
\end{equation}
Performing the Kramer's-Moyal expansion to first order we find this gives
\begin{equation}
\{ g^{ab}(y), \ch(y)\rho \} +  \Tr(h^{\ag \bg} \lin_{\bg}^{\dag} \lin_{\ag} \{ g^{ab}, \cqstate\}) + \dots 
\end{equation}
and smeared over an observable $A$ reads
\begin{equation}
\int Dg D \pi \{ A, g^{ab}(y) \} ( \ch(y)\rho +  \Tr(h^{\ag \bg} \lin_{\bg}^{\dag} \lin_{\ag} \cqstate)) + \dots
\end{equation}
and looks exactly like what we might expect to get as a Hamiltonian constraint in the classical limit. 
\item The third line is the part coming from the Poisson bracket of the constraint with $\mathcal{\bar{L}}$, which, as mentioned, we expect to cancel with the $\mathcal{R}$ terms (in the fourth and fifth lines) due to the fact that $\bar{\mathcal{C}} ^a$ transforms like a momentum and $\bar{\mathcal{L}}$ like a Hamiltonian. This does not appear to happen here. Taking the trace over the quantum system, the third and fourth lines of equation \eqref{eq: fullham} combine to give
\begin{equation}
\begin{split}
& \int d^3 x NM_a  \bigg[ \Tr\left[\bar{\mathcal{L}}(\{ g^{ab}, \cqstate\} ) - \bar{\mathcal{L}}(\cm_b  \{ g^{ab}, \cqstate \}) + (2 \parts_{JH}^{\ag \bg} - 2\parts_{NH}^{\ag \bg})(\cqstate) L_{\bg}^{\dag} L_{\ag}) \right] \\
& + \Tr{}{\left[(2\parts_{HJ}^{\ag \bg}-  2\parts_{HN}^{\ag \bg}) (\cqstate) \lin_{\bg}^{\dag} \lin_{\ag}  \right] } \bigg].
\label{eq: intbyparts}
\end{split}
\end{equation} 
If we perform a Kramers Moyal expansion to first order we see that the first term in \eqref{eq: intbyparts} vanishes, so that we are left with the final term, $\Tr{}{\left[(2\parts_{HJ}^{\ag \bg}-  2\parts_{HN}^{\ag \bg}) (\cqstate) \lin_{\bg}^{\dag} \lin_{\ag}  \right] }$. Recalling the definition of $R_{HJ}, R_{HN}$ in equation \eqref{eq: partsdef}, as well as the form of $C_N$ in \eqref{eq: definitions}, we see that the $R_{HN}$ term cancels with the zeroth moment of the $R_{HJ}$ term and we are left with the first moment of the  $2\parts_{HJ}^{\ag \bg}$ term alone. Explicitly, $\parts_{HJ}^{\ag \bg}$ is written
\begin{equation}
\begin{split}
\parts_{HJ}^{\ag \bg}= & \int d^3x N\big[ W^{\phi \phi} D_b(M_a\con_H^{ab}\cqstate ) - W^{\pi \pi}  D_b (M_a\con_H^{ab}\cqstate)  + (D_b(M_a\con_H^{af}\cqstate ) W^{eb}\\
& + D_b(M_a\con_H^{ae}\cqstate ) W^{bf} -D_b(M_a\con_H^{ab}\cqstate) W^{ef})  + M_a C_H^{ab}  L_{\ag} D_b(W^{\ag \bg}\cqstate) L_{\bg} \big],
\label{eq: maybe another constraint}
\end{split}
\end{equation}
and one can verify if the first moments of the Kramers-Moyal expansion are to give GR then we know that this will not identically be zero. One might hope that one could include it in the Hamiltonian constraint, but since it comes with different smearing functions to the (would be) Hamiltonian constraint in the first two lines of \eqref{eq: fullham}, we must either place a restriction on the lapse and shift, or equation \eqref{eq: maybe another constraint} must be imposed as a separate constraint. This does not necessarily over-constrain the system, since it is akin to a constraint on the density matrix, and the CQ state has more freedom. For example, one can constrain the various moments of its distribution without reducing the number of degrees of freedom of the theory.

We also see that there are offending terms of the form $D_b(C_H^{ab} \cqstate)$, these are interesting since we do not get these in pure GR; the matter part of the momentum constraint contains no metric degrees of freedom. It is worth noting that we get these violating terms even in classical analogues of the CQ theory. To be precise, one can ask the question: can one have a Markovian master equation  on the phase space of GR which contains noise and is gauge invariant? For example, one can study a Fokker-Planck type equation which is linear in the lapse and shift and apply the same arguments as outlined in section \ref{sec: dirac} to derive a momentum constraint, which must then be preserved in time. In doing so, we still find violating terms of the form $D_b(C^{ab} \cqstate)$. We hope to discuss this further in simple classical toy models, where one can study this violation more explicitly.

\item The rest of the terms in \eqref{eq: fullham} are purely CQ terms with no classical analogue. They come with different smearing functions to those in the (would be) Hamiltonian constraint in the first two lines of equation \eqref{eq: fullham} and so would either need to be imposed as a separate constraint, itself preserved in time, or we could impose restrictions on the choice of lapse and shift; we do not check what this gives here. Given the form of the remaining terms, it seems that they render the constraint to be non-satisfiable without restrictions on the lapse and shift. One notes that all of the terms come from the $C_J, C_{JN}$ and $ C_N$ parts of the momentum constraint, which in turn come from the Jump, and no-event parts of the evolution equation when calculating the $[\mathcal{L}[N], \mathcal{L}[M]]$ commutation, i.e, from the consideration of the gauge algebra under two CQ jumps. This is, perhaps, where one might have expected the current demands for gauge invariance to break down. One might hope that if there was no no-event term in the evolution equation then the expression for the remaining terms in \eqref{eq: fullham} is greatly simplified. This is the case for the class of continuous classical-quantum dynamics introduced in \cite{UCLPawula}, and the constraints for such dynamics will be presented in a future work using the methodology introduced here \cite{oppenheim2021constraints}. More generally, it would seem that elimination of these terms will require us to weaken the assumptions outlined in Section \ref{sec: scalarfield}. We therefore leave it open as a possibility that one could find a different form of gauge invariance to that presented in this paper.
\end{itemize}

To summarize, although we have found a sensible looking momentum constraint, when looking at its conservation in time we get equation \eqref{eq: fullham} which looks like it will either require a restriction on the choice of lapse and shift, or additional constraints to be satisfied. We have broken down \eqref{eq: fullham} into multiple terms with different interpretations. In particular, we find several different terms which come with different smearing functions, meaning we do not find a single constraint, but we must either impose multiple constraints or put a restriction on the lapse and shift so that either the smearing functions vanish, or are equivalent to one another. One could argue that we can keep on imposing more and more constraints à la Dirac and hope that some way down the line we end up with a weakly closing algebra defining so that our assumptions regarding gauge invariance are respected. However, one must be extremely careful when doing this not to over-constrain the system; we would like a theory with 2 degrees of freedom per space-time point. Understanding what this means precisely in our formalism is somewhat convoluted. Take for example the pure gravity momentum constraint as derived in equation \eqref{eq: classcon}, $ \langle \{ g^{ab}, A \}, \mathcal{H}_b \rho \rangle \approx 0$. As mentioned, we can take this in its weakest form; it needs to vanish when smeared over observables, as opposed to arbitrary smearing functions. As a consequence, we view the constraint as a constraint on the moments of any allowed distribution $\rho$ and it is very difficult to quantify how many degrees of freedom one is eliminating when imposing constraints on the moments only. However, when looking at the evolution of the weak form of the constraint, one finds that the constraint under action of the momentum gauge generator gives  $
\langle \{  \{ g^{ab}, A \}, \mathcal{H}_c \}, \mathcal{H}_b  \rho \rangle =0
$, whilst under the action of the Hamiltonian gauge generator gives back a Hamiltonian constraint and a term $
\langle \{  \{ g^{ab}, A \}, \mathcal{H}_c \}, \mathcal{H}  \rho \rangle =0
$ which is like the momentum constraint but smeared over a different operator. This is not necessarily zero if  $ \langle \{ g^{ab}, A \}, \mathcal{H}_b \rho \rangle \approx 0$ vanishes for some subset of possible $A$. As discussed, we can keep on applying gauge transformations to get more and more constraints on the moments of the distribution $\rho$, but it would appear that one is forced eventually to impose a stronger version of the constraint: instead of viewing \eqref{eq: classcon} as a constraint on the moments of the distribution, we could interpret it as a constraint on the state space, which must hold for any smearing function $A$. In imposing the constraint on the state space, one can define a sensible notion of degrees of freedom, at least in the classical case. For example, in a classical stochastic theory the probability distribution is a map from the phase space $\Gamma$ to the unit interval, $\rho: \Gamma \to [0,1]$. We define the space of possible configurations of probability distributions which satisfy the constraint as $\mathcal{S}_{\mathcal{C}}$. The dimension of $\mathcal{S}_{\mathcal{C}}$ is then a quantifier of the number of degrees of freedom in the theory. For example, in pure GR one can pick a basis of $\mathcal{S}_{\mathcal{C}}$ to be given by $\delta(g- \bar{g}, \pi- \bar{\pi})$, where both $\bar{g}$ and $\bar{\pi}$ satisfy the Hamiltonian and momentum constraints. If we want a theory with 2 degrees of freedom per space-time point we must then impose 4 independent local constraints; imposing more than this is over-constraining the system, yielding less than two degrees of freedom per space-time point. However, constraints which restrict the probability distribution itself, (for example, restrict its mean and variance), do not lead to over-constraining.

\section{Discussion}\label{sec: discussion}

In this work, we have presented a methodology to study the gauge invariance of a class of Markovian CQ theories of gravity linear in the lapse and shift $N, N^a$ which reproduce the dynamics of GR in the classical limit. We have here presented the algebra for a class of realisations of the full theory, which we call the ``discrete class''. The theory could be regarded as fundamental, or as an effective theory of quantum gravity in the classical limit of the gravitational degrees of freedom. Since the theory is not unitary, and isn't derived from a local action, we have derived the constraints on the level of the equations of motion. The theory is invariant under spatial diffeomorphisms. We then demand that the theory be invariant under an arbitrary choice of $N$ and $N^a$, and from this, we are able to derive the constraints of the theory, including an analog of the momentum and Hamiltonian constraint. The momentum constraint arises as a condition required for the theory to be invariant under the choice of lapse function, while the Hamiltonian constrain arises from demanding that the momentum constraint be preserved in time. Unlike classical GR, the constraints do not correspond to a constraint surface of the phase space, but rather are an operator equation acting on the state. The operators in question are operators acting on the Hilbert space, along with functional derivatives with respect to phase-space degrees of freedom (or more correctly, classical finite difference operators). This is reminiscent of approaches in quantum gravity where the super-Hamiltonian and super-momentum operator are applied to the state and one constrains the wave-function to be annihilated by these operators. However, here, we don't assume that the classical constraints are just promoted to operators, but rather, derive them from the symmetry we impose. Invariance under the choice of lapse and shift, is not necessarily equivalent to diffeomorphism invariance, but is natural in a phase-space formulation of the theory.

Here, we have asked for the largest possible gauge group on phase space we could imagine to hold, while still being consistent with General Relativity. However, such invariance is perhaps too strong, as a number of theories such as Horava gravity, shape dynamics and the unimodular theory, fix the lapse and shift, usually with a global fixing. For the class of models studies here, subject to the particular simplifying assumptions made in Section \ref{sec: scalarfield}, 
we find that such a global condition on the lapse and shift might be necessary, in order for the constraints to close. The gauge group of the theory would then be smaller than the gauge group of general relativity. One could for example be satisfied with spatial diffeomorphisms alone, and impose that $N$ must be spatially constant. In that case, the constraint algebroid would close since the smearing functions always appear with divergences acting on them. However, this would be too strong a restriction, since one typically needs to retain greater freedom in choosing the lapse and shift, to avoid the evolution becoming singular. One hopes a weaker one can be found. Alternatively, one hopes that weakening one or more of the Assumptions \ref{as:linear}-\ref{as:pure}  on the realisations might allow for the constraint algebroid to close without any restrictions on $N$ and $N^a$. In particular,  the ``continuous class'' of realisations \cite{UCLPawula} are obtained when Assumption \ref{as:1} is weakened \cite{oppenheim2021constraints}.

Another possibility is that if one thinks of this theory as the classical limit of quantum gravity, then we should lift the Markovian assumption, since by taking the classical limit, we are in effect throwing away quantum information which could act as a memory for the evolution. In this case, one hopes the methodology  sheds light on a possible theory of quantum gravity. Indeed the algebra shares some features of the quantum algebra, since the matter fields are quantised, while the classical nature of the gravity part allows for a much more tractable set of calculations. Along the way, one sees that a number of conceptual issues that prove difficult to resolve in quantum gravity, also occur in theories where one has a probability density over gravitational degrees of freedom.

{\bf Acknowledgements}
We would like thank Joan Camps, Henrique Gomes, Philipp Hoehn, Andrea Russo and Lee Smolin for valuable discussions. JO is supported by an EPSRC Established Career Fellowship, and a Royal Society Wolfson Merit Award and Z.W.D.~acknowledges financial support from EPSRC. This research was supported by the National Science Foundation under Grant No. NSF PHY11-25915 and by the Simons Foundation {\it It from Qubit} Network.  

\appendix

\section{Realisations for \texorpdfstring{$W^{\ag\bg}(z|z';x)$}{real}}\label{sec: realisations}

In the main text we considered realisations to be consistent with Assumptions \ref{as:linear}-\ref{as:pure} but otherwise tried to be as general as possible in our calculations of the constraint algebra without considering a particular realisation of $W^{\ag\bg}(z|z';x)$. In this section we take up the task of constructing such realizations. The form of the functions $W^{\ag\bg}(z|z';x)$ are very constrained by the requirement that \eqref{eq: block1} defines a positive matrix, by spatial diffeomorphism invariance and by the requirement that their first moment reproduces general relativity in the classical limit (this first moment is itself dictated by spatial diffeomorphism invariance). For this section, we shall explicitly use a spatial regulator $\epsilon(x-y)$, since we were only able to construct local realizations which are UV divergent, and require regularisation. We suspect that the class of classical-quantum master equations introduced in \cite{UCLPawula} could be used to construct realizations with better UV properties but this is beyond the scope of the current discussion.

Some realisations for $W^{\alpha \beta}(x)$ were given in \cite{oppenheim_post-quantum_2018}. A key ingredient is to note that the couplings $h^{\alpha \beta}$ in equation \eqref{eq: h for scalar field} are all positive. As a consequence, for the scalar field, we can take the positive functionals
\begin{align}\label{eq: realizationphi}
    W^{\phi\phi}(x)&=\frac{1}{2 \tau}\sqrt{g}e^{\tau \int dy \epsilon(x-y) g^{-1/2}(y)\{\sqrt{g}(y),\cdot\}}\nonumber\\
    &=     \frac{1}{2 \tau}\sqrt{g}e^{\frac{\tau}{2} \int dy \epsilon(x-y) g^{ij}\frac{\delta}{\delta \pi^{ij}}}
    \nonumber\\
    &= \frac{\sqrt{g}}{2 \tau}
    +\frac{1}{4}\int dy\epsilon(x-y)\sqrt{g}g^{ij}\frac{\delta}{\delta \pi^{ij}(y)}  \nonumber\\
    & \s \s +\frac{1}{8}\tau\int dy_1 dy_2
    \epsilon(y_1-x)\epsilon(y_2-x)\sqrt{g}g^{ij}g^{kl}\frac{\delta^2}{\delta \pi^{ij}(y_1)\delta \pi^{kl}(y_2)}+\cdots
    \end{align}
    \begin{align}\label{eq: realizationpi}
    W^{\pi\pi}(x)&=\frac{1}{2 \tau}g^{-1/2}e^{\tau\int dy \epsilon(x-y) \sqrt{g}(y)\{g^{-1/2}(y),\cdot\}}
    \nonumber\\
    &=\frac{1}{2 \tau}g^{-1/2}e^{-\frac{\tau}{2}\int dy \epsilon(x-y)g^{ij}\frac{\delta}{\delta \pi^{ij}} }
    \nonumber\\
      &= \frac{{g}^{-1/2}}{2 \tau}
    -    \frac{1}{4}\int dy\epsilon(x-y){g}^{-1/2}g^{ij}\frac{\delta}{\delta \pi^{ij}(y)}  \nonumber\\
    & \s \s +\frac{1}{8}\tau\int dy_1 dy_2 \epsilon(y_1-x)\epsilon(y_2-x){g}^{-1/2}g^{ij}g^{kl}\frac{\delta^2}{\delta \pi^{ij}(y_1)\delta \pi^{kl}(y_2)} + \cdots ,
\end{align}
and we see that if we expand the exponent, then the zeroth moment gives the Lindbladian couplings, the first moments give the general Relativistic matter couplings and the second moments give the diffusion terms. In both \eqref{eq: realizationphi} and \eqref{eq: realizationpi} we would like to take $\epsilon(x-y)$ to be arbitrarily peaked after computing any quantity and removing UV divergences. If it is left finite, then one gets finite quantities, but violates cluster decomposition over some short range \cite{bps,OR-intrinsic} and $\epsilon(x-y)$ becomes a parameter of the theory as in fundamental decoherence models rather than just being a method to regularise the theory. This is of course not a problem if viewing the dynamics as effective. 

The realisation for $W^{ab}(x)$ is trickier because $g^{ab}$ is not a scalar nor square matrix that can be exponentiated. We could take
\begin{align}
      W^{ab}(x)&=\frac{1}{4\tau}\sqrt{g(x)}g_{ac}
      \exp[
    \tau \int dy \epsilon(x-y)M
   ]^c_b
    +\frac{1}{4\tau}\sqrt{g(x)}g_{bc}  \exp[
    \tau \int dy \epsilon(x-y)M
   ]^c_a
    \nonumber\\
    &=     \frac{1}{2\tau}\sqrt{g(x)}g^{ab} + \frac{1}{2}\int dy\epsilon(x-y)\{\sqrt{g}g^{ab},\cdot\}+\cdots
\end{align}
with $M$ being the square matrix with components $g^{-1/2}g^{ci}\{\sqrt{g}g_{ib},\cdot\}$ and $\exp\left[M\right]$ being the matrix
exponential with components $\exp \left[M\right]^c_b$. However, when acted on a scalar functional, this doesn't produce a tensor. 
One option is to perform a canonical transformation, and consider the phase space to be given by the tetrad (or {\it vierbein}) and its conjugate coordinates, by writing the metric in terms of the tetrad fields $E^a_i$
\begin{align}
    g^{ab}=E^a_iE^b_j\eta^{ij},
\end{align}
with $\eta^{ij}$ being the flat spatial metric. We can
then use the matrix exponential on the square matrices $E$ with components $E^a_i$ to give
\begin{align}
    :W^{ab}(x):&=\frac{1}{4\tau}\sqrt{g}g^{ac}\exp\left[
    \tau \int dy \epsilon(x-y)E^{-1}\{E,\cdot\}
    \right]_c^b
    +  \nonumber\\
    & \s \s + \frac{1}{4\tau}\sqrt{g}g^{bc}\exp\left[
    \tau \int dy \epsilon(x-y)E^{-1}\{E,\cdot\}
    \right]_c^a,
\end{align}
where we have written the components of the exponential of the matrix $M$ as $\exp[M]^\alpha_i$, and the Poisson bracket is with respect to the tetrad fields and conjugate momenta \cite{ashtekarvar}. One then has to impose an additional constraint to preserve $\eta^{ij}$. This is a general construction for any $h^{\ag\bg}$ which is decomposable as a sum of scalars or square matrices, whether they be tetrads $E^a_i$ or any other square matrix $M^\alpha_i$. Such a construction can also be used for a stochastic pure gravity term.

Thus far, we have taken $N(x,t)$, $N^a(x,t)$ to depend on $x,t$, and we've then allowed the smearing functions for gauge transformations to depend explicitly on the metric $g_{ab}$. However, one could consider the case where one allows $N$ and $N^a$ to depend explicitly on the metric in the master equation. In classical GR, one can even take the lapse and shift to depend on the matter fields. Here, the choice of time parameterisation is a classical notion, so having it depend on quantum fields would make little sense. We would also argue that while the lapse and shift could depend on $g_{ab}$ it should not depend on $\pi^{ab}$, since this is a fluctuating quantity which branches in each trajectory. In either case, normalization of probability requires that if the lapse and shift is a functional of phase space variables, it needs to appear inside the Poisson bracket and the higher derivative terms, as can be seen from equation \eqref{eq: expansion}. We can then consider realisations of the form
\begin{align}
      W^{\ag\bg}[N]&=\int\frac{\lambda^{ij}}{4\tau}NM^{\alpha}_iM^{\gamma}_j(x)\exp\left[
    \tau \int dy \epsilon(x-y)M^{-1}N^{-1/2}\{N^{1/2}M,\cdot\}
    \right]_\gamma^\beta
    \nonumber\\
    &+
     \int\frac{\lambda^{ij}}{4\tau}NM^\beta_iM_j^\gamma(x)\exp\left[
    \tau \int dy \epsilon(x-y)M^{-1/2}N^{-1/2}\{N^{1/2}M^{1/2},\cdot\}
    \right]_\gamma^\alpha
\end{align}
for a decomposition of a Hamiltonian in terms of square matrices $M$, we can write 
\begin{equation}
    \lapse(x)=M^\mu_{\sigma} M^\nu_{\gamma}(x)\lambda^{\sigma \gamma} L^\dagger_\nu(x) L_\mu(x),
\end{equation} 
and similarly for $W^{\ag \bg}[\vec{N}]$, here $\mu$ runs over $\mathbb{I}, \alpha$. This construction can also be used for the pure gravity evolution, which is encoded in the identity component. If one were to then require that this be equivalent to one in which the lapse and shift depend solely on $x,t$, one finds both additional constraints which are reminiscent of the CQ Hamiltonian and momentum constraint. In fact, such a requirement in the case of classical GR leads to exactly the Hamiltonian and momentum constraint. However, in the CQ case one finds constraints which have a dependence on the lapse and shift, and one therefore should regard them as additionally constraining the gauge group of the theory.

\bibliographystyle{JHEP}
\bibliography{refgrav2}
\end{document}